\documentclass[aps,prb,twocolumn,superscriptaddress,showpacs,longbibliography]{revtex4-1}

\usepackage{amsmath,amssymb,graphicx,subfigure}

\usepackage[utf8]{inputenc}
\usepackage[T1]{fontenc}
\usepackage{xcolor}
\usepackage[normalem]{ulem}
\usepackage{multirow}

\IfFileExists{newtxtext.sty}
   {\usepackage{newtxtext,newtxmath}}
   {\IfFileExists{stix.sty}
      {\usepackage{stix}}
      {\IfFileExists{mathptmx.sty}
      {\usepackage{mathptmx}}{} } }

\usepackage{textcomp}

\usepackage{bm}

\IfFileExists{siunitx.sty}{\usepackage{booktabs,siunitx}}{}

\pdfoutput=1
\usepackage{color}
\definecolor{LinkColor}{rgb}{0.256,0.439,0.588}
\usepackage{hyperref}
\hypersetup{
   colorlinks=true,
   citecolor=LinkColor,
   linkcolor=LinkColor,
   urlcolor=LinkColor
}

\newcommand{\be}{\begin{equation}}
\newcommand{\ee}{\end{equation}}
\newcommand{\bea}{\begin{eqnarray}}
\newcommand{\eea}{\end{eqnarray}}

\usepackage{pifont}

\begin{document}

\title{Classification of phase transitions and intertwined orders in crystallographic point groups}

\author{Heqiu Li}
\affiliation{Department of Physics, University of Michigan, Ann Arbor, MI 48109, USA}

\date{\today}

\begin{abstract}

Symmetry provides important insight in understanding the nature of phase transitions. In the presence of crystalline symmetries, new phenomena in phase transition can emerge, such as intertwined orders and emergent symmetries. In this work, we present a group-theoretical diagnosis of these phenomena in 32 crystallographic point groups. For each group, we classify the possible combinations of intertwined order and analyze different symmetry-broken phases. We find that the symmetries before and after the phase transition uniquely determine the irreducible representation (irrep) of the order parameter and the allowed form of Ginzburg-Landau free energy, from which the nature of phase transition can be inferred, including the order of phase transition and the existence of intermediate phases and emergent symmetries. Our finding will be helpful for the experimental diagnosis of order parameters from symmetry breaking.

\end{abstract}
\pacs{pacs numbers}
\maketitle

\section{Introduction}

An important insight from modern physics is that different phases in nature can be classified by their symmetries. The Landau theory \cite{Landau1937} characterizes phase transitions by spontaneous symmetry breaking. During the phase transition, the order parameter acquires a finite value, and the symmetries under which the order parameter is not invariant will be broken. The broken symmetry can be internal symmetry, e.g., the global U(1) symmetry in superconductors, or it can also be spatial and time-reversal symmetry. The classical and quantum nematic phases \cite{nematicbook1993,chaikin1995principles,Kivelson1998,Fradkin1999,Oganesyan2001} break rotation symmetry. The familiar ferromagnetic and ferroelectric phases break time-reversal and space-inversion symmetry respectively, and both symmetries are broken in the multiferroics \cite{Kimura2003,Cheong2007,Cheong2018} and ferro-toroidal phase \cite{VanAken2007,Spaldin_2008,Zimmermann2014}. The phase with ferro-rotational order \cite{Gopalan2011,Johnson2012,Terada2015,Jin2020} does not break time-reversal or space-inversion symmetry, but it can be characterized by other crystalline symmetries. Phase transitions lead to many observable physical phenomena, and the broken symmetry can be detected in various experiments such as crystallography and optical experiments \cite{hahn1983international,Castellan2002,DEVRIES1986193,Harter295,B511119F,Matlack2014}.

The microscopic origin of phase transitions is generally diversified and sensitive to system details, but symmetries can enforce strong and universal constraints on phase transitions. For example, the symmetry breaking pattern determines whether a phase transition can be second-order, e.g., nematic transitions in liquid crystals have to be first-order by symmetry~\cite{chaikin1995principles}. Symmetry also determines whether an intermediate phase is present, e.g., the transition from isotropic phase to smectic phase has an intermediate nematic phase and may go through two separate transitions. Many novel phenomena in phase transitions are related to symmetry, such as the intertwined orders \cite{Fradkin2015,Fernandes2019} and emergent symmetries \cite{Blankschtein1984,Isakov2003,Gazit2018}. Because symmetries can usually be determined conveniently by experiments, studying the relation between symmetries and phase transitions can provide valuable insight in the diagnosis of phase transitions. The analysis that relates symmetry to phase transition is achieved by group theory, and several works have studied phase transitions through a group-theoretical approach \cite{Hlinka2016,Watanabe2018,Erb2018,Norman2020}. In the presence of symmetry, the order parameter transforms as irreducible representations (irrep) of the symmetry group, which imposes a constraint on the form of Ginzburg-Landau free energy and provides implications on the fate of the phase transition. In this work, we establish a diagnosis of phase transitions from symmetry breaking patterns in the 32 crystallographic point groups, such that once we are given the symmetries before and after some phase transition, we will know what is the generic form that the Ginzburg-Landau free energy can take, what irreducible representation (irrep) does the order parameter have and whether this phase transition has an intermediate phase. This diagnosis is shown in Fig. \ref{gpattern}. Based on this symmetry analysis, we can classify the following emergent phenomena related to the interplay between crystalline symmetries and order parameters in each symmetry group:

1. The distinct symmetry-broken phases. When the dimension of irrep is higher than 1, the order parameter in a given irrep can lead to multiple distinct low-temperature phases with different symmetries depending on the microscopic details of the free energy. For example, in a cubic crystal with $O_h$ symmetry, the free energy of the 3D Heisenberg ferromagnetic acquires an additional anisotropic term $f_A=\lambda (\psi_x^4+\psi_y^4+\psi_z^4)$ and the ferromagnetic phase possess a preferred direction of magnetization along either <111> or <100> direction depending on whether $\lambda$ is positive or negative, leading to different low-temperature phases with distinct subgroup symmetry.

2. Intertwined orders \cite{Fradkin2015,Fernandes2019}. Different order parameters may intertwine with each other such that they are no longer independent. The presence of two order parameters in different irreps $A$ and $B$ can induce another order parameter in irrep $C$ through a term $\psi_A\psi_B\psi_C$ in free energy, provided this triple product is allowed by symmetry. In this case, any two of them can induce the third one. We construct a table for each point group the possible combinations of irreps that the intertwined order can occur.

3. Emergent symmetries \cite{Blankschtein1984,Isakov2003,Gazit2018}. Near a phase transition, a new symmetry that does not belong to the original system may emerge in the Ginzburg-Landau free, because the terms that break this symmetry are irrelevant under renormalization group. An example is shown later in the $E_u$ irrep of $O_h$ group, where there is an emergent SO(3) symmetry and the system is in the 3D Heisenberg universality class. The symmetry breaking pattern can tell whether the system has an emergent symmetry near the critical point. We are mainly interested in emergent continuous symmetries (ECS) that can change the universality class of the critical point~\cite{Isakov2003}.

\begin{figure}
\includegraphics[width=2.8 in]{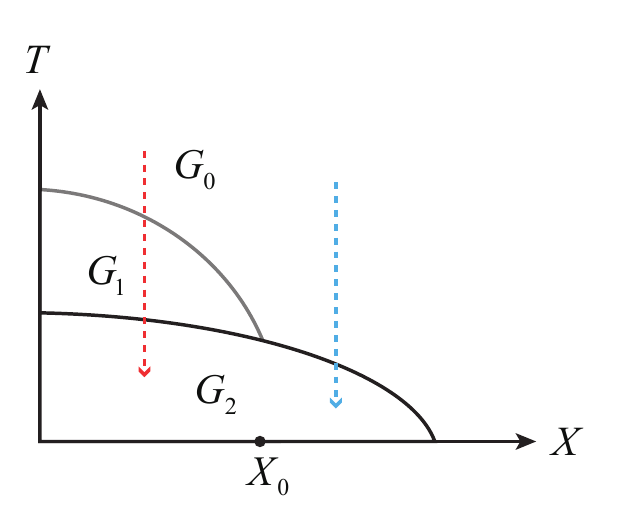}
\caption{ Schematic phase diagram for a phase transition sequence. $T$ is temperature and $X$ is an arbitrary control parameter. As the temperature is lowered, the system initially at the high temperature phase with symmetry group $G_0$ may go through two transitions to $G_1$ and then to $G_2$ following the red arrow, or go through one first-order phase transition to $G_2$ directly following the blue arrow.   }
\label{figspt}
\end{figure}

4. Phase transition sequence. If the free energy has multiple anisotropic terms favoring different symmetry-broken phases, the system can go through a sequence of phase transitions when temperature is lowered. For example, suppose a cubic system is initially in a high temperature phase with vanishing order parameter, and its free energy has an $O(\psi^3)$ term favoring the order parameter along <111> direction and an $O(\psi^4)$ term favoring <100> direction. As we lower the temperature, there are two possible ways of phase transitions. One possibility is that the system first has a phase transition to <111> phase, then as the temperature is further lowered, the $O(\psi^4)$ term becomes comparable to $O(\psi^3)$ term and the system undergoes another phase transition to <100> phase. The other possibility is that the system may go through a large first-order phase transition so that it directly jumps from the high temperature phase to <100> phase. A schematic phase diagram is shown in Fig. \ref{figspt}, where $T$ is temperature and $X$ is an arbitrary control parameter. $G_0$, $G_1$ and $G_2$ are symmetry groups of different phases. The two possibilities of phase transitions above correspond to the evolution along the red and blue arrows respectively.

Our classification focus on order parameters that have a uniform spatial distribution with momentum $\mathbf{q}=0$ and can be characterized by point group symmetry. The order parameters with finite momentum such as charge density wave and FFLO pairing order are deferred to future work. This classification is achieved by explicitly writing down the extra anisotropic terms in the free energy that are compatible with crystalline symmetry. We demonstrate our method through an example of the classification in systems with $O_h$ symmetry.

\section{Emergent phenomena in phase transitions}
\label{phenomena}

We demonstrate the emergent phenomena in phase transitions such as the phase transition sequence, the intertwined order and the distinct symmetry-broken phases through the analysis of order parameter that transforms as irrep $E_g$ in the $O_h$ group. The analysis for the other irreps are shown in Section \ref{classifyOh}.

The order parameter with irrep $E_g$ can arise during a nematic phase transition. We begin by writing down the Ginzburg-Landau free energy that is compatible with the symmetry requirement. Denote the two real components of order parameter by $\psi_1$ and $\psi_2$. For irrep $E_g$, $\psi_1$ and $\psi_2$ transform like $2z^2-x^2-y^2$ and $\sqrt{3}(x^2-y^2)$ respectively. Define $\psi=\psi_1+i\psi_2=|\psi|e^{i\phi}$ so that $|\psi|=\sqrt{\psi_1^2+\psi_2^2}$ is the magnitude of the order parameter. To write down the free energy density $f$, we need to find combinations of $\psi_1$ and $\psi_2$ that transform as $A_{1g}$. The traditional isotropic terms that are $\phi$-independent are always allowed:
\be
f_{iso}=\frac{r}{2}|\psi|^2+\frac{u}{4}|\psi|^4+g_0|\psi|^6+...
\label{fiso}
\ee
To find out the other allowed terms, we consider the effect of each generator of $O_h$ on $\psi$. The absolute value $|\psi|$ is invariant under all group elements and the only change is in the phase $\phi$. For a fourfold rotation along z direction, $\psi_1\sim 2z^2-x^2-y^2$ does not change but $\psi_2 \sim\sqrt{3}(x^2-y^2)$ changes sign, therefore the effect of $C_{4z}$ is $\phi\rightarrow -\phi$. Similar analysis can be carried out for other generators, which gives
\bea
&&C_{4z}:\phi\rightarrow -\phi \nonumber\\
&&C_3:\phi\rightarrow \phi+\frac{2\pi}{3} \nonumber\\
&&{C}_2':\phi\rightarrow  -\phi \nonumber\\
&&I:\phi\rightarrow \phi
\eea
Here $C_3$ is along [111] direction, ${C}_2'$ is along [110] direction and $I$ is space inversion. The requirement that every term in the free energy density should be invariant under these transformations gives:
\be
f=f_{iso}-w|\psi|^3\cos3\phi+g|\psi|^6\cos6\phi,
\label{f36}
\ee
where $w$ and $g$ are free parameters. The total free energy is obtained by the integral of $f$ over 3D space $F=\int f d^3x$. The existence of these anisotropic terms is because the crystal breaks the $O(3)$ symmetry to $O_h$.

The free energy density in the form of Eq.\eqref{f36} will appear multiple times in our classification scheme, which is worthwhile for a detailed study. The phase transition described by Eq.\eqref{f36} is in sharp contrast to the case with only isotropic terms. First we omit the sixth order terms for simplicity, and the free energy density becomes
\be
f=\frac{r}{2}|\psi|^2-w|\psi|^3\cos3\phi+\frac{u}{4}|\psi|^4
\ee
If $w>0$, the free energy will be minimized at $\phi=\frac{2\pi}{3}n$ and if $w<0$ the free energy will be minimized at $\phi=\frac{2\pi}{3}n+\pi$. In either case the system has a first order phase transition when $r=\frac{2w^2}{u}$, and $|\psi|$ jumps discontinuously from 0 to $\frac{2|w|}{u}$. Note that the low temperature phase has preferred orientation of order parameter with $\phi=\frac{\pi}{3}n$, and the symmetry is spontaneously broken to $D_{4h}$. For the $\phi=0$ phase when $w>0$ and the $\phi=\pi$ phase when $w<0$, the remaining $D_{4h}$ group consists of a $D_{2h}$ subgroup with $C_{2x},C_{2y},C_{2z}$ and an additional $C_{4z}$ symmetry. For the other phases with $\phi=\pm\frac{2\pi}{3}$ and $\pm\frac{\pi}{3}$, the remaining $D_{4h}$ group consist of the same $D_{2h}$ subgroup but with the additional $C_4$ symmetry along either x or y. When these $D_{4h}$ groups for different low temperature phases are viewed as subgroups of the original $O_h$ group, these subgroups are conjugate to each other by a $C_3$ rotation. This is necessary because $C_3$ is the operator that relates different low temperature phases by sending $\phi$ to $\phi+\frac{2\pi}{3}$.

\begin{figure}
\includegraphics[width=3.4 in]{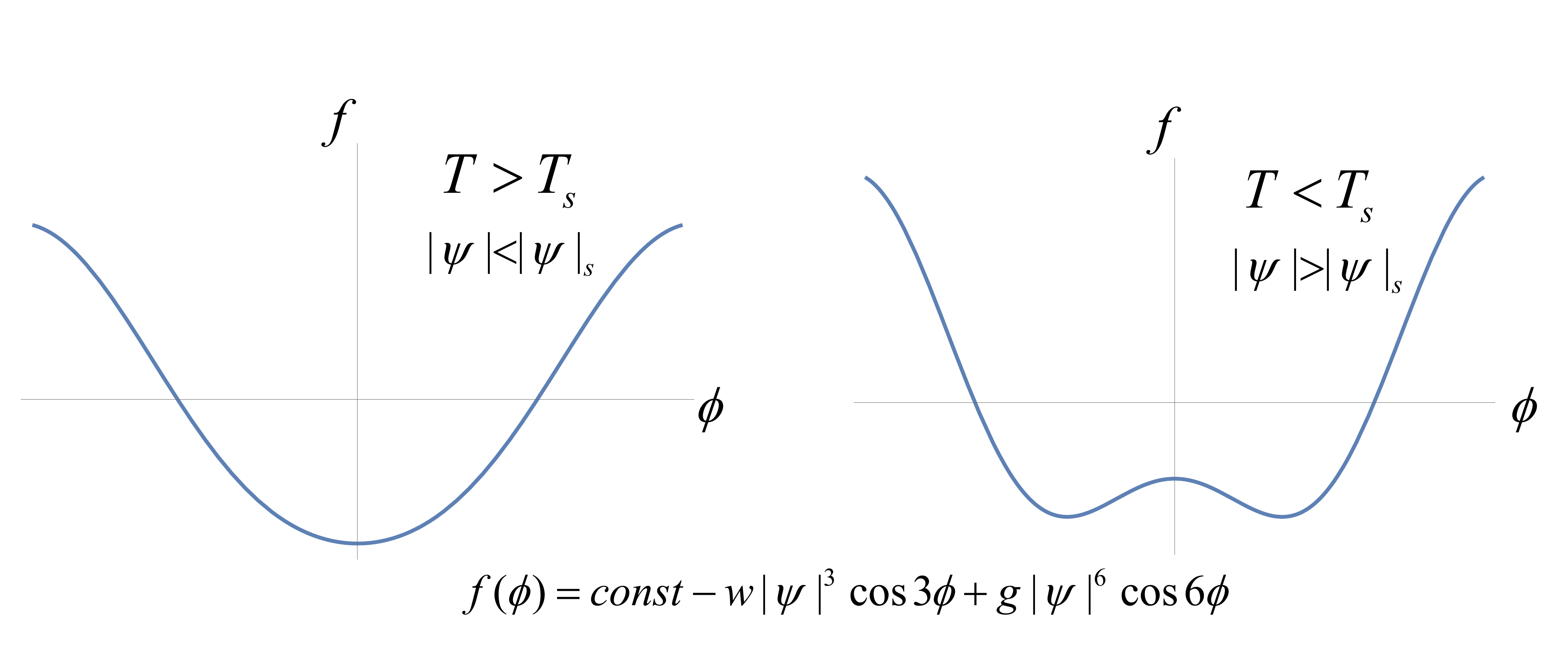}
\caption{ The free energy density as a function of $\phi$ in Eq.\eqref{f36} for $w>0,g>0$ near $\phi=0$. There is a second order transition at $T_s$ when $|\psi|=|\psi|_s=(\frac{|w|}{4g})^{1/3}$.  }
\label{transition}
\end{figure}

There is a phase transition sequence in which the $D_{4h}$ symmetry can be further broken when we take into account the $O(|\psi|^6)$ terms. After the transition to $D_{4h}$, as the temperature is further lowered, in Eq.\eqref{f36} $|\psi|$ may increase so that $w\sim g|\psi|^3$. If $g<0$, then the $O(|\psi|^6)$ terms favor the same ground state as the $O(|\psi|^3)$ term, and no new phase transition occurs. However, if $g>0$, with sufficiently large $|\psi|$ the ground state can deviate from $\phi=\frac{\pi}{3}n$ and go through another phase transition, breaking the $D_{4h}$ symmetry to $D_{2h}$. It turns out that the critical condition is met when $|\psi|=|\psi|_s=(\frac{|w|}{4g})^{1/3}$. We denote the corresponding temperature as $T_s$. The free energy density as a function of $\phi$ near one minima is shown in Fig. \ref{transition}. The transition at $T_s$ is second order, with a spontaneous symmetry breaking from $D_{4h}$ to $D_{2h}$. The transition $O_h-D_{4h}-D_{2h}$ is an example of a phase transition sequence. It has a phase diagram similar to Fig. \ref{figspt}. When $|\psi|_s=(\frac{|w|}{4g})^{1/3}>\frac{2|w|}{u}$, the high temperature phase goes through two phase transitions (first to $D_{4h}$ and then to $D_{2h}$) as we decrease the temperature, corresponding to the red arrow in Fig. \ref{figspt}. When $|\psi|_s<\frac{2|w|}{u}$, because the order parameter during the first-order transition jumps from 0 to $\frac{2|w|}{u}$, the system jumps directly to the phase with $D_{2h}$ symmetry through a first-order phase transition, corresponding to the blue arrow.

\begin{figure}
\includegraphics[width=2.4 in]{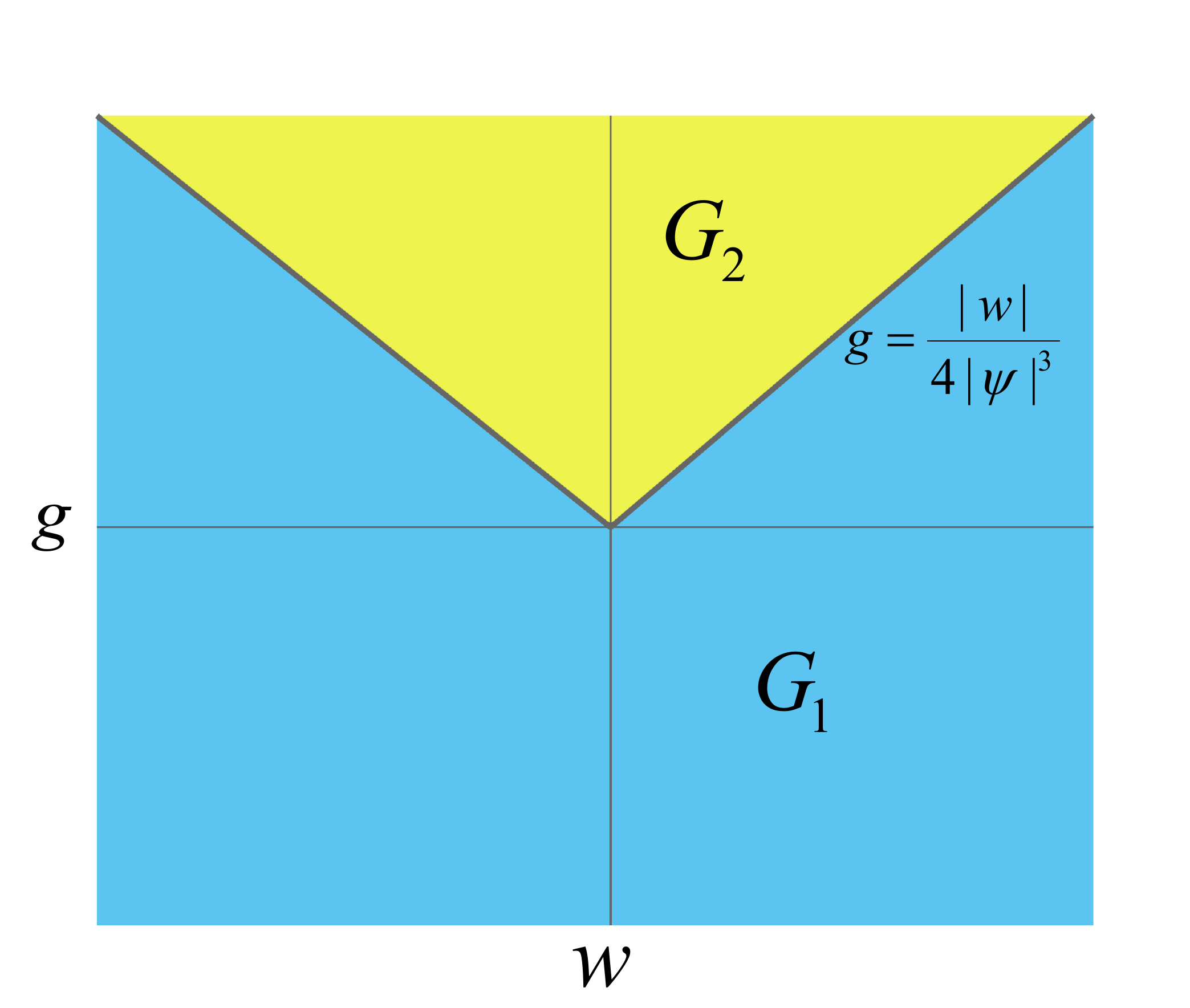}
\caption{ The symmetry of low temperature phases in Eq.\eqref{f36}. There are two phases denoted by regions with different colors, and there is a second-order phase transition between them. Phase 1 has $g<\frac{|w|}{4|\psi|^3}$ with symmetry group $G_1$. If $w>0$, phase 1 has $\phi=\frac{2\pi n}{3}$; if $w<0$, phase 1 has $\phi=\frac{2\pi n}{3}+\pi$. Phase 2 has $g>\frac{|w|}{4|\psi|^3}$ with symmetry group $G_2$. In phase 2, $\phi$ is not fixed to any high-symmetry value.  }
\label{phase36}
\end{figure}

The phase diagram of systems described by Eq.\eqref{f36} is summarized as follows. After the first order transition at $r=\frac{2w^2}{u}$, $|\psi|$ acquires a finite value and the $O_h$ symmetry is broken to its subgroups. The low temperature phases are classified into two regions in Fig. \ref{phase36}. When $|\psi|<|\psi|_s$, or equivalently $g<\frac{|w|}{4|\psi|^3}$, the system is in region 1 with $\phi=\frac{2\pi}{3}n$ for $w>0$ and $\phi=\frac{2\pi}{3}n+\pi$ for $w<0$. The symmetry group in this case is $G_1$ isomorphic to $D_{4h}$. When $g>\frac{|w|}{4|\psi|^3}$, the system is in region 2 with $\phi$ away from high symmetry values, and the symmetry is $G_2$ isomorphic to $D_{2h}$.

The intertwined order in group $O_h$ can be found by checking the triple product of irreps $\Gamma_1\otimes\Gamma_2\otimes\Gamma_3$. If the trivial irrep $A_{1g}$ is contained in the product, them the order parameters in two of $\Gamma_1,\Gamma_2$ and $\Gamma_3$ can generate the third one. Note that it is required that $\Gamma_1,\Gamma_2$ and $\Gamma_3$ are different irreps, which is distinct from the case of induced secondary order parameter with a term like $\Gamma_1\Gamma_1\Gamma_2$ in free energy. For example, in group $O_h$ the triple product $A_{1u}\otimes E_u\otimes E_g=A_{1g}\oplus A_{2g}\oplus E_g$ contains the trivial irrep $A_{1g}$, them there is a term $\sim \psi_{A_{1u}}\psi_{E_u}\psi_{E_g}$ allowed in the free energy and these three order parameters are intertwined. The list of the combination of intertwined order parameters in group $O_h$ is shown in table \ref{table_Oh}.


\begin{table*}
\centering
\begin{tabular}{ |c||c|c|c|c|c|  }
 \hline
 \multicolumn{6}{|c|}{$O_h$} \\
 \hline
  & $f_A$ &  Symmetry of low temperature phases & FOT & PTS &ECS   \\
 \hline
 $E_g$& $-w|\psi|^3\cos3\phi+g|\psi|^6\cos6\phi$  & Fig. \ref{phase36}: $G_1=D_{4h};\ G_2=D_{2h}(3C_{2})$   &Y&Y&N\\
 \hline
 $E_u$& $g|\psi|^6\cos6\phi$  & $g<0:G_1=D_{4};\ g>0:G_2=D_{2d}(S_{4},C_{2})$   &N&N&Y\\
 \hline
 $T_{1g}$& $\lambda(\psi_{x}^4+\psi_{y}^4+\psi_{z}^4)$  & $\lambda>0:G_1=C_{3i};\ \lambda<0:G_2=C_{4i}$   &N&N&N\\
 \hline
 $T_{1u}$& $\lambda(\psi_{x}^4+\psi_{y}^4+\psi_{z}^4)$  & $\lambda>0:G_1=C_{3v};\ \lambda<0:G_2=C_{4v}$   &N&N&N\\
 \hline
 $T_{2g}$& $-w\psi_{xy}\psi_{yz}\psi_{xz}+\lambda(\psi_{xy}^4+\psi_{yz}^4+\psi_{xz}^4)$  & Fig. \ref{phase34}: $G_1=D_{3d};\ G_2=D_{2h}(C_{2},C_{2}')$   &Y&Y&N\\
 \hline
 $T_{2u}$& $\lambda(\psi_{xy}^4+\psi_{yz}^4+\psi_{xz}^4)$  & $\lambda>0:G_1=D_{3};\ \lambda<0:G_2=D_{2d}(S_{4},C_{2}')$   &N&N&N\\
 \hline
 IO &\multicolumn{5}{c|}{$ \{ET_1T_2 \},\{A_2T_1T_2\},T_{1u}T_{2u}T_{2g},E_uT_{2u}T_{2g},E_uT_{1u}T_{2u},A_{2u}E_uE_g ,A_{1u}T_{2u}T_{2g},A_{1u}T_{1u}T_{1g},A_{1u}E_uE_g,A_{1u}A_{2u}A_{2g}   $}  \\
 \hline
\end{tabular}
\caption{Table for phase transition in systems with $O_h$ symmetry. The notations are introduced in Section \ref{notation}. $f_A$ is the anisotropic part of free energy. FOT, PTS, ECS and IO are shorthand for first order transition, phase transition sequence, emergent continuous symmetry and intertwined orders respectively.   }
\label{table_Oh}
\end{table*}

\section{Notations for the tables and figures}\label{notation}

Before we move on to discuss the other irreps and symmetry groups, we summarize the main features in group $O_h$ in table \ref{table_Oh}, and the same notations in this table will be used for other groups as well. In the table, $f_A$ refers to the anisotropic parts of free energy. FOT, PTS, ECS and IO are shorthand for first order transition, phase transition sequence, emergent continuous symmetry and intertwined order respectively, and their existence is denoted by Y/N (yes/no). Some of the group generators are added to specify the group. For example, $D_{2h}(3C_{2})$ means this $D_{2h}$ group contains three $C_2$ axes along x,y,z directions; whereas $D_{2h}(C_{2},C_2')$ means the group has one $C_2$ operator conjugate to $C_{2z}$, and two other $C_2$ operators conjugate to the twofold rotation along $\hat x+\hat y$ direction.

The intertwined orders in each group are shown in the table as well. Each triple product in the row labeled by "IO" represents a combination of intertwined orders. The notation $\{ET_1T_2 \}$ is a shorthand for the collection of four triple products with different arrangements of parity: $E_gT_{1g}T_{2g}$, $E_gT_{1u}T_{2u}$, $E_uT_{1g}T_{2u}$ and $E_uT_{1u}T_{2g}$.

\begin{figure*}
\includegraphics[width=6.8 in]{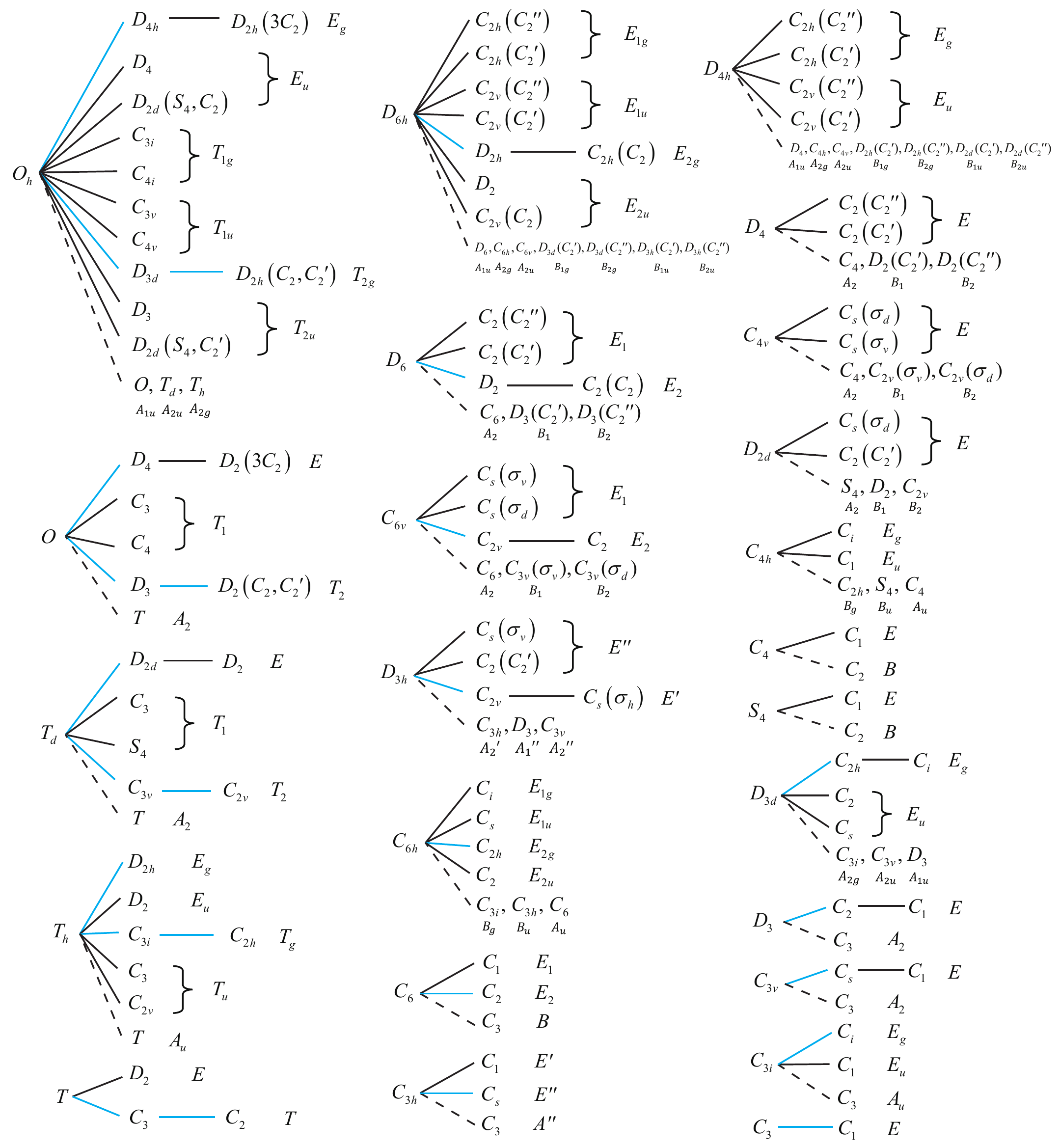}
\caption{ Different possibilities of symmetry breaking during phase transitions for each group. Each figure begins with the symmetry group of the high temperature phase. The blue (black) lines refer to first-(second-)order phase transition. The dashed lines represent phase transitions with order parameter in one-dimensional representations. Phase transition sequences are shown as connected lines. }
\label{gpattern}
\end{figure*}

The different possibilities of symmetry breaking during phase transitions for each group are summarized in Fig. \ref{gpattern}. Each figure begins with the group of the high temperature phase, followed by lines linking to symmetry-broken low temperature phases. The blue lines represent first-order transitions and the black lines represent second-order transitions. The dashed lines represent phase transitions with order parameters that transform under 1D irreps, and these transitions are second-order as well. Each possibility of symmetry breaking uniquely determines the irrep of order parameter as shown in the figure. The phase transition sequences are shown as connected lines, such as the $O_h-D_{3d}-D_{2h}$ sequence. Each phase transition sequence indicates a phase diagram similar to Fig. \ref{figspt}, where the high temperature phase $G_0$ can either go through a phase transition to $G_1$ and then to $G_2$ when temperature is lowered, or by a large first-order phase transition from $G_0$ to $G_2$ directly.

To utilize our results in diagnosis of phase transition, one can start from the symmetries before and after the phase transition which are obtained experimentally, then refer to Fig. \ref{gpattern} to identify the irrep of order parameter. The information on the phase transition can then be found by referring to the corresponding table for the given irrep. This includes the anisotropic part of the Ginzburg-Landau free energy $f_A$, the parameter range for each possible symmetry-broken low temperature phases, and whether this transition can be first-order (FOT), whether there is a phase transition sequence (PTS) or emergent continuous symmetry (ECS). If the irrep of order parameter is one-dimensional according to Fig. \ref{gpattern}, then the free energy is fully isotropic with $f_A=0$, and the entries for FOT, PTS and ECS are all negative.

\section{Phase transition in systems with point group symmetry}

In this section we apply our analysis to all the crystallographic point groups and classify the emergent phenomena in phase transitions. Any order parameter can be split into components such that each component transforms under one irrep, hence we only need to consider one irrep at a time. Our main focus is on the order parameters that transform as high-dimensional irreps, because the the order parameters in these irreps have multiple components, allowing distinct symmetry-broken phases. The phase transitions related to 1D irreps are also shown in Fig. \ref{gpattern}. We do not need to consider the double group irreps because order parameters always transform under single group irreps. The irreps for each group can be found in, e.g., Ref~\onlinecite{Dresselhaus:2010aa,pointgroup}.

\subsection{Systems with $O_h$ symmetry}
\label{classifyOh}

The high-dimensional irreps in group $O_h$ are the 2D irreps $E_g$, $E_u$ and 3D irreps $T_{1g},T_{1u},T_{2g},T_{2u}$. The analysis for irrep $E_g$ has been carried out in Section \ref{phenomena}. Now we focus on the other irreps. The features of systems with $O_h$ symmetry are summarized in table \ref{table_Oh}.

\subsubsection{Representation $E_u$}

The order parameter in $E_u$ has two components $\psi_1\sim(2z^2-x^2-y^2)A_{1u}$ and $\psi_2\sim\sqrt{3}(x^2-y^2)A_{1u}$, where $A_{1u}$ refers to a pseudo scalar function that is invariant under all proper symmetries in $O_h$ but changes sign under inversion. Define $\psi=\psi_1+i\psi_2=|\psi|e^{i\phi}$ as before. Inversion reverses the sign of both $\psi_1$ and $\psi_2$ and sends $\phi$ to $\phi+\pi$. Therefore, the $\cos3\phi$ term in Eq.\eqref{f36} is no longer allowed and the free energy becomes
\be
f=f_{iso}+g|\psi|^6\cos6\phi
\ee
This system has an emergent symmetry SO(3), because the anisotropic term appears only in sixth order, which is irrelevant in renormalization group for 3D space. This is an example of emergent continuous symmetry (ECS). The phase transition is still second order, but the low temperature phase depends on the sign of $g$. If $g<0$ then $\phi=\frac{\pi n}{3}$ is preferred, and the symmetry spontaneously breaks from $O_h$ to $G_1=D_4$. For the phase with $\phi=0$, this $D_4$ group contains $C_{4z}$ and $C_{2x}$. If $g>0$ then the system prefers $\phi=\frac{\pi n}{3}+\frac{\pi}{6}$, and the symmetry reduces from $O_h$ to $G_2=D_{2d}$. For $\phi=\frac{\pi}{2}$, this $D_{2d}$ group contains $S_{4}$ rotoinversion along z direction and $C_2$ along x,y,z directions.

\subsubsection{Representation $T_{2g}$}

The order parameter has three components $\psi_{xy},\psi_{yz}$ and $\psi_{xz}$ that transform as $xy,yz$ and $xz$ respectively. Define $|\psi|=\sqrt{\psi_{xy}^2+\psi_{yz}^2+\psi_{xz}^2}$. The general form of free energy is
\be
f=f_{iso}-w\psi_{xy}\psi_{yz}\psi_{xz}+\lambda(\psi_{xy}^4+\psi_{yz}^4+\psi_{xz}^4)+O(|\psi|^6)
\label{f34}
\ee
We neglect $O(|\psi|^6)$ terms because they do not lead to new phases, unlike in Eq.\eqref{f36} where $O(|\psi|^6)$ terms are needed to further break the symmetry. There could also be an anisotropic term like $\psi_{xy}^2\psi_{xz}^2+\psi_{xy}^2\psi_{yz}^2+\psi_{yz}^2\psi_{xz}^2$ at fourth order, but this term is proportional to $|\psi|^4-(\psi_{xy}^4+\psi_{yz}^4+\psi_{xz}^4)$, therefore it can be converted to the $\lambda(\psi_{xy}^4+\psi_{yz}^4+\psi_{xz}^4)$ term in the free energy.

Consider the case with $\lambda=0$ first. If $w>0$, for a fixed $|\psi|$, the free energy is minimized when $(\psi_{xy},\psi_{yz},\psi_{xz})\sim(1,1,1),(-1,-1,1),(-1,1,-1),(1,-1,-1)$. If $w<0$, the minima occurs when $(\psi_{xy},\psi_{yz},\psi_{xz})\sim(-1,-1,-1),(1,-1,1),(1,1,-1),(-1,1,1)$. In both cases the system has a first order transition to a phase with $|\psi_{xy}|=|\psi_{yz}|=|\psi_{xz}|$ at low temperature, and the symmetry group is broken to $D_{3d}$. For the phase with $(\psi_{xy},\psi_{yz},\psi_{xz})\sim(1,1,1)$, this $D_{3d}$ group contains $C_3$ along [111] direction and one of the $C_2$ is along $\hat x-\hat y$ direction.

\begin{figure}
\includegraphics[width=2.4 in]{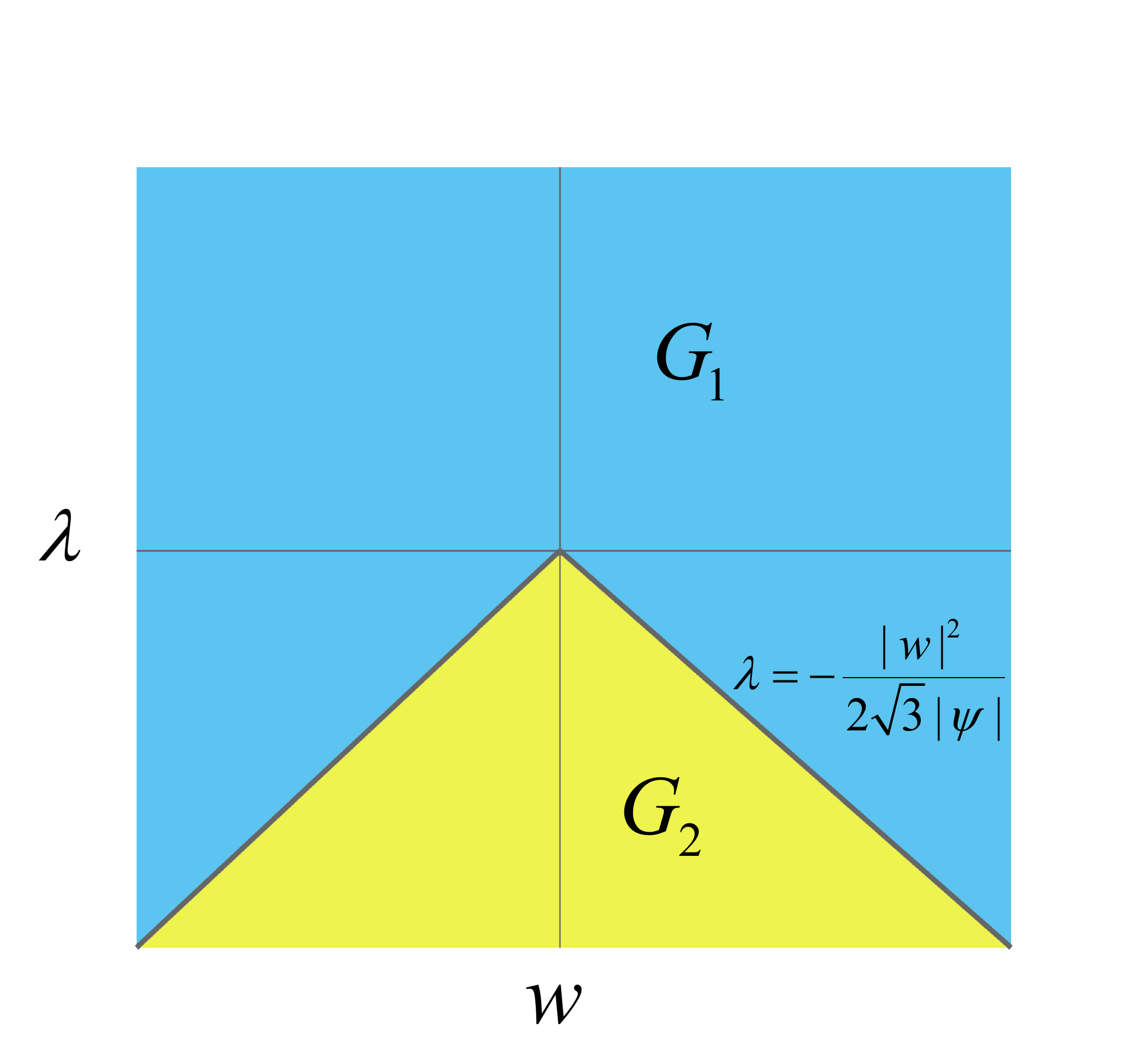}
\caption{ The symmetry of low temperature phases in Eq.\eqref{f34}. There are two phases denoted by different colors, and there is a first-order phase transition between them. Phase 1 has $\lambda>-\frac{|w|}{2\sqrt{3}|\psi|}$ with symmetry group $G_1$. In phase 1 the three components of the order parameter has the same magnitude. Phase 2 has $\lambda<-\frac{|w|}{2\sqrt{3}|\psi|}$ with symmetry group $G_2$. In phase 2 only one of the three components of the order parameter is nonzero. }
\label{phase34}
\end{figure}

The presence of $\lambda$ in Eq.\eqref{f34} can lead to a phase transition sequence with a phase diagram similar to Fig. \ref{figspt}. If $\lambda>0$, then the the $\lambda$ term is minimized when $|\psi_{xy}|=|\psi_{yz}|=|\psi_{xz}|$ as well, and there is no new phase transition. We denote this phase with $|\psi_{xy}|=|\psi_{yz}|=|\psi_{xz}|$ as phase 1. If $\lambda<0$, then the $\lambda$ term favors a phase where only one of $\psi_{xy},\psi_{yz},\psi_{xz}$ is nonzero and the other two vanishes, which we denote as phase 2. Minimization of free energy in Eq.\eqref{f34} at a given $|\psi|$ for general $\lambda$ and $w$ shows that the minimum of free energy always occurs in either phase 1 or phase 2. When $\lambda>-\frac{|w|}{2\sqrt{3}|\psi|}$ the system is in phase 1 with symmetry group $G_1$ isomorphic to $D_{3d}$, and  When $\lambda<-\frac{|w|}{2\sqrt{3}|\psi|}$ the system is in phase 2 with symmetry group $G_2$ isomorphic to $D_{2h}$, as shown in Fig. \ref{phase34}. If the nonzero component in phase 2 is $\psi_{xy}$, the $C_2$ in this symmetry group $D_{2h}$ are along z, $\hat x+\hat y$ and $\hat x-\hat y$ directions. Note that this $D_{2h}$ group is distinct from the $D_{2h}$ discussed previously for irrep $E_g$, whose $C_2$ axes are along coordinate axes.


\subsubsection{Representation $T_{2u}$}

The three components of order parameter $\psi_{xy},\psi_{yz}$ and $\psi_{xz}$ in $T_{2u}$ transform as $xy A_{1u}, yzA_{1u}$ and $xzA_{1u}$ respectively, where $A_{1u}$ is a rotational invariant pseudo scalar function. The $O(|\psi|^3)$ term is no longer allowed in free energy, which implies
\be
f=f_{iso}+\lambda(\psi_{xy}^4+\psi_{yz}^4+\psi_{xz}^4)+O(|\psi|^6)
\ee
The phase transition is second order, and the low temperature phase is determined by the sign of $\lambda$. If $\lambda>0$, the systems favors phase 1 with $|\psi_{xy}|=|\psi_{yz}|=|\psi_{xz}|$, and the symmetry is broken to $G_1=D_3$. For $(\psi_{xy},\psi_{yz},\psi_{xz})\sim(1,1,1)$, the $D_3$ group has $C_3$ along [111] direction and one of the $C_2$ rotation is along $\hat x-\hat y$. If $\lambda<0$, the system favors phase 2 with only one of $\psi_{xy},\psi_{yz},\psi_{xz}$ being nonzero, and the symmetry is reduced from $O_h$ to $G_2=D_{2d}$. If the nonzero component is $\psi_{xy}$, then this $D_{2d}$ group has $S_4$ along z direction and $C_2$ along $\hat x \pm\hat y$ directions. Note that this $D_{2d}$ group is distinct from the $D_{2d}$ obtained previously for $E_u$ representation.

\subsubsection{Representation $T_{1g}$ and $T_{1u}$}

The $T_{1g}$ and $T_{1u}$ order parameters transform as axial vectors and polar vectors respectively. The free energy for both systems have the same form as $T_{2u}$:
\be
f=f_{iso}+\lambda(\psi_{x}^4+\psi_{y}^4+\psi_{z}^4)+O(|\psi|^6)
\ee
The analysis is identical to that in $T_{2u}$. The phase transition is second order. For $T_{1g}$, if $\lambda>0$, the low temperature phase has symmetry $C_{3i}$, and if $\lambda<0$ the symmetry is $C_{4i}$, where the $C_3$ and $C_4$ axes are parallel to the component of order parameter.

For $T_{1u}$, the $\lambda>0$ phase has symmetry $C_{3v}$. For $(\psi_{x},\psi_{y},\psi_{z})\sim(1,1,1)$, the $C_3$ in the $C_{3v}$ group is along [111] direction and one of the mirror planes is perpendicular to $\hat x-\hat y$. The $\lambda<0$ phase has symmetry $C_{4v}$, where the $C_4$ axis is parallel to the order parameter, and the mirror planes pass through the $C_4$ axis.

\subsection{Systems with $O$ or $T_d$ symmetry}

The groups $O$ and $T_d$ are isomorphic and many features are shared by them. In particular, isomorphic groups have the same irreps, and have the same form of free energy. Therefore, the existence/absence of first order transition, phase transition sequence, emergent symmetry and intertwined order are the same for isomorphic groups. The isomorphism between these groups maps the symmetry of the low temperature phase of one group to that of the other group.

The properties of the irreps in group $O$ is the same as that for the corresponding parity-even irreps in group $O_h$. Therefore, for irrep $E$ of group $O$, the free energy has the same form as Eq.\eqref{f36}, with the existence of first order transition and phase transition sequence. With the same condition as in Fig. \ref{phase36}, the low symmetry phases for group $O$ are $G_1=D_4$ and $G_2=D_2(3C_{2})$, which are obtained by removing space inversion in the corresponding $G_1$ and $G_2$ in the case of $O_h$. For irrep $T_{2}$, the free energy is the same as Eq.\eqref{f34}, and the low temperature phases become $G_1=D_{3}$ and $G_2=D_2(C_{2},C_{2}')$. For irrep $T_1$, we have $G_1=C_3$ and $G_2=C_4$.

The analysis of $T_d$ is carried out by the group isomorphism. The isomorphism maps $C_4$ and $C_2'$ in $O$ to $S_4$ and $\sigma_d$ in $T_d$, where $\sigma_d$ is a mirror plane along diagonal direction, therefore the irrep $E$ in $T_d$ has $G_1=D_{2d}$ and $G_2=D_2$. Similarly, irrep $T_{1}$ has $G_1=C_3$ and $G_2=S_4$. For irrep $T_2$, the components of order parameter $\psi_{yz},\psi_{xz},\psi_{xy}$ in group $O$ are mapped respectively to $\psi_x,\psi_y,\psi_z$ in $T_d$, which transform as $x,y,z$ respectively. Irrep $T_{2}$ has $G_1=C_{3v}$ and $G_2=C_{2v}$. The table for $O$ and $T_d$ are shown in table \ref{table_OTd}.

\begin{table*}
\centering
\begin{tabular}{ |c|c||c|c|c|c|c|  }
 \hline
 \multicolumn{7}{|c|}{$O$ and $T_d$} \\
 \hline
  & & $f_A$ &  Symmetry of low temperature phases & FOT & PTS &ECS   \\
 \hline
\multirow{3}{*}{$O$}& $E$& $-w|\psi|^3\cos3\phi+g|\psi|^6\cos6\phi$  & Fig. \ref{phase36}: $G_1=D_{4};\ G_2=D_{2}(3C_{2})$   &Y&Y&N\\
 \cline{2-7}
 &$T_{1}$& $\lambda(\psi_{x}^4+\psi_{y}^4+\psi_{z}^4)$  & $\lambda>0:G_1=C_{3};\ \lambda<0:G_2=C_{4}$   &N&N&N\\
 \cline{2-7}
 &$T_{2}$& $-w\psi_{xy}\psi_{yz}\psi_{xz}+\lambda(\psi_{xy}^4+\psi_{yz}^4+\psi_{xz}^4)$  & Fig. \ref{phase34}: $G_1=D_{3};\ G_2=D_{2}(C_2,C_2')$ &Y&Y&N\\
 \hline
 \multirow{3}*{$T_d$}& $E$& $-w|\psi|^3\cos3\phi+g|\psi|^6\cos6\phi$  & Fig. \ref{phase36}: $G_1=D_{2d};\ G_2=D_{2}$   &Y&Y&N\\
 \cline{2-7}
 &$T_{1}$& $\lambda(\psi_{x}^4+\psi_{y}^4+\psi_{z}^4)$  & $\lambda>0:G_1=C_{3};\ \lambda<0:G_2=S_{4}$   &N&N&N\\
 \cline{2-7}
 &$T_{2}$& $-w\psi_{x}\psi_{y}\psi_{z}+\lambda(\psi_{x}^4+\psi_{y}^4+\psi_{z}^4)$  & Fig. \ref{phase34}: $G_1=C_{3v};\ G_2=C_{2v}$ &Y&Y&N\\
 \hline
 \multicolumn{2}{|c||}{IO} &\multicolumn{5}{c|}{$ ET_1T_2 ,\ A_2T_1T_2  $}  \\
 \hline
\end{tabular}
\caption{Table for isomorphic groups $O$ and $T_d$. The notations are introduced in Section \ref{notation}.}
\label{table_OTd}
\end{table*}

\subsection{Systems with $T$ symmetry}

Group $T$ has high dimension irrep $E$ and $T$. The partners in irrep $E$ can be chosen to transform as $\psi\sim (2z^2-x^2-y^2)+i\sqrt{3}(x^2-y^2)=|\psi|e^{i\phi}$. The action of $C_3$ along [111] direction generates $\phi\rightarrow\phi+\frac{2\pi}{3}$, and $C_2$ does not change $\phi$. The general form of free energy for irrep $E$ is
\be
f=f_{iso}-w|\psi|^3\cos(3\phi+\delta)+g|\psi|^6\cos(6\phi+\delta')
\label{fdel}
\ee
At the low temperature phase, $|\psi|$ acquires a finite value through a first order transition and breaks the symmetry from $T$ to $D_2$. Unlike Eq.\eqref{f36} for $E_g$ in group $O_h$, the presence of the arbitrary constants $\delta$ and $\delta'$ in Eq.\eqref{fdel} does not fix $\phi$ to any high symmetry values, which means the symmetry of this ground state cannot be further broken by changing the value of $\phi$. Therefore, the low temperature phase for order parameter in irrep $E$ has only one type of symmetry $G=D_2$, contrary to the previous examples where there are distinct phases with symmetry $G_1$ and $G_2$. It implies that there is no phase transition sequence.

For irrep $T$, the components of order parameters $\psi_x,\psi_y,\psi_z$ transforms as $x,y,z$ respectively. The general free energy for irrep $T$ is
\be
f=f_{iso}-w\psi_x\psi_y\psi_z+\lambda(\psi_x^4+\psi_y^4+\psi_z^4)
\ee
This is the same as Eq.\eqref{f34}, and the same analysis can be applied here. There is a first order phase transition, and distinct low temperature phases $G_1$ and $G_2$ that occur under the conditions in Fig. \ref{phase34}. Here $G_1=C_3$ and $G_2=C_2$. The transition between the low temperature phase with symmetry $G_1$ and $G_2$ is a first-order phase transition, by the same analysis under Eq.\eqref{f34}.

\begin{table*}
\centering
\begin{tabular}{ |c||c|c|c|c|c|  }
 \hline
 \multicolumn{6}{|c|}{$T$} \\
 \hline
  & $f_A$ &  Symmetry of low temperature phases & FOT & PTS &ECS   \\
 \hline
 $E$& $-w|\psi|^3\cos(3\phi+\delta)+g|\psi|^6\cos(6\phi+\delta')$  & $G=D_2$   &Y&N&N\\
 \hline
 $T$& $-w\psi_x\psi_y\psi_z+\lambda(\psi_x^4+\psi_y^4+\psi_z^4)$  & Fig. \ref{phase34}: $G_1=C_3; \ G_2=C_2 $   &Y&Y&N\\
 \hline
 IO &\multicolumn{5}{c|}{None}  \\
 \hline
\end{tabular}
\caption{Table for group $T$. The notations are introduced in Section \ref{notation}.   }
\label{table_T}
\end{table*}

\subsection{Systems with $T_h$ symmetry}

The phase transition in the parity-even irreps of $T_h$ follows the same analysis as group $T$, with the only change that the symmetry of the low temperature phases in $T_h$ are obtained from the symmetry groups of $T$ by a direct product with group $C_i$.

The parity-odd irreps have different features because the third order term is no longer allowed in free energy. For irrep $E_u$ the free energy is
\be
f=f_{iso}+g|\psi|^6\cos(6\phi+\delta)
\ee
Because the free energy is isotropic up to fourth order in $|\psi|$, the system has an emergent continuous symmetry. Contrary to the situation in the irrep $E_u$ in group $O_h$, here there is only one type of symmetry $G=D_2$ for the low temperature phase no matter $g$ is positive or negative.

For irrep $T_u$, the free energy is
\be
f=f_{iso}+\lambda(\psi_x^4+\psi_y^4+\psi_z^4)
\ee
It has the same form as irrep $T_{1u}$ in group $O_h$. The phase transition is second order. If $\lambda>0$, the low temperature phase has symmetry $G_1=C_3$ with $|\psi_x|=|\psi_y|=|\psi_z|$. If $\lambda<0$, the low temperature has symmetry $G_2=C_{2v}$.

\begin{table*}
\centering
\begin{tabular}{ |c||c|c|c|c|c|  }
 \hline
 \multicolumn{6}{|c|}{$T_h$} \\
 \hline
  & $f_A$ &  Symmetry of low temperature phases & FOT & PTS &ECS   \\
 \hline
 $E_g$& $-w|\psi|^3\cos(3\phi+\delta)+g|\psi|^6\cos(6\phi+\delta')$  & $G=D_{2h}$   &Y&N&N\\
 \hline
 $E_u$& $g|\psi|^6\cos(6\phi+\delta)$  & $G=D_{2}$   &N&N&Y\\
 \hline
 $T_g$& $-w\psi_x\psi_y\psi_z+\lambda(\psi_x^4+\psi_y^4+\psi_z^4)$  & Fig. \ref{phase34}: $G_1=C_{3i}; \ G_2=C_{2h} $   &Y&Y&N\\
 \hline
 $T_u$& $\lambda(\psi_x^4+\psi_y^4+\psi_z^4)$  & $\lambda>0: G_1=C_3; \ \lambda<0:G_2=C_{2v} $   &N&N&N\\
 \hline
 IO &\multicolumn{5}{c|}{$ T_uT_gE_u,\ T_uT_gA_u $}  \\
 \hline
\end{tabular}
\caption{Table for group $T_h$. The notations are introduced in Section \ref{notation}.   }
\label{table_Th}
\end{table*}

\subsection{Systems with $D_6,C_{6v}$ or $D_{3h}$ symmetry}

The groups $D_6,C_{6v}$ and $D_{3h}$ are isomorphic. We choose the convention that the group $D_6$ to have its primary axis along z, one of the three $C_2'$ rotations is $C_{2y}$, and one of the three $C_2''$ rotations is $C_{2x}$. The group $C_{6v}$ has primary axis along z and one of the $\sigma_v$ mirror pass through axes x and z. The group $C_{3h}$ has horizontal mirror $\sigma_h$, and one of its $C_2'$ axis is along x. Under the group isomorphism, symmetry $C_6$ in group $D_6$ is mapped to $C_6$ in $C_{6v}$ and $S_3=C_3\sigma_h$ in $D_{3h}$, $C_2'$ in $D_6$ is mapped to $\sigma_d$ in $C_{6v}$ and $C_2'$ in $D_{3h}$, and $C_2''$ in $D_{6}$ is mapped to $\sigma_v$ in $C_{6v}$ and $\sigma_v$ in $D_{3h}$. There are two 2D irreps $E_1$ and $E_2$, with $E_1$ ($E_2$) in group $D_6$ being the irrep whose $C_6$ character is $+1$ ($-1$).

$E_1$: We choose the order parameter in irrep $E_1$ in groups $D_6$ and $C_{6v}$ to transform as $\psi=|\psi|e^{i\phi}\sim x+iy$, and for group $C_{3h}$ $\psi=|\psi|e^{i\phi}\sim(x+iy)z$. For group $D_6$, the action of group elements on $\phi$ is:
\bea
C_{6z}&:&\phi\rightarrow\phi+\pi/3 \nonumber\\
C_{2x}&:&\phi\rightarrow -\phi \nonumber\\
C_{2y}&:&\phi\rightarrow \pi-\phi
\eea
The action on the other groups are the same if we replace these symmetry operators by the corresponding ones under group isomorphism. These transformation properties restricts the free energy to be $\phi$-independent until sixth order:
\be
f=f_{iso}+g|\psi|^6\cos6\phi
\ee
Therefore, the system with an order parameter in irrep $E_1$ has emergent continuous symmetry. The phase transition is second order. If $g<0$, the low temperature phase has $\phi=\frac{n\pi}{3}$, and the symmetry is broken to subgroup $G_1=C_2(C_2'')$ for group $D_6$, $C_s(\sigma_v)$ for group $C_{6v}$, and $C_s(\sigma_v)$ for group $D_{3h}$. If $g>0$, the low temperature phase has $\phi=\frac{n\pi}{3}+\frac{\pi}{6}$, and the symmetry is broken to subgroup $G_2=C_2(C_2')$ for group $D_6$, $C_s(\sigma_d)$ for group $C_{6v}$, and $C_2(C_2')$ for group $D_{3h}$.

$E_2$: We choose the order parameter in $E_2$ to transform as $\psi=|\psi|e^{i\phi}\sim(x^3-3xy^2)(x+iy)$ for $D_6$ and $C_{6v}$, and $\psi=|\psi|e^{i\phi}\sim(x+iy)$ for $D_{3h}$. For group $D_6$ the action of group elements on $\phi$ is:
\bea
C_{6z}&:&\phi\rightarrow\phi-2\pi/3 \nonumber\\
C_{2x}&:&\phi\rightarrow -\phi \nonumber\\
C_{2y}&:&\phi\rightarrow -\phi
\eea
Comparing with $E_1$, the term of cubic power on $\psi$ is allowed in the free energy for $E_2$:
\be
f=f_{iso}-w|\psi|^3\cos3\psi+g|\psi|^6\cos6\phi
\ee
It has the same form as Eq.\eqref{f36}. Following the analysis under Eq.\eqref{f36}, the phase transition is first order, with a successive second order phase transition when temperature is further lowered. There are two distinct low temperature phases with symmetry $G_1$ and $G_2$, as shown in Fig. \ref{phase36}. The symmetry of phase 1 is reduced to $G_1=D_2(C_{2}'')$ for group $D_6$, and $G_1=C_{2v}$ for $C_{6v}$ and $D_{3h}$. The symmetry of phase 2 is reduced to $G_2=C_2(C_{2z})$ for group $D_6$, and $G_2=C_{2}$ for $C_{6v}$, and $G_2=C_s(\sigma_h)$ for $D_{3h}$.

\begin{table*}
\centering
\begin{tabular}{ |c|c||c|c|c|c|c|  }
 \hline
 \multicolumn{7}{|c|}{$D_6$, $C_{6v}$ and $D_{3h}$} \\
 \hline
  & & $f_A$ &  Symmetry of low temperature phases & FOT & PTS &ECS   \\
 \hline
 \multirow{2}*{$D_6$}& $E_1$& $g|\psi|^6\cos6\phi$  & $g<0:G_1=C_{2}(C_{2}'');\ g>0:G_2=C_{2}(C_{2}')$    &N&N&Y\\
 \cline{2-7}
 &$E_2$& $-w|\psi|^3\cos3\phi+g|\psi|^6\cos6\phi$  & Fig. \ref{phase36}: $G_1=D_{2};\ G_2=C_{2}(C_{2})$   &Y&Y&N\\
 \hline
 \multirow{2}*{$C_{6v}$}& $E_1$& $g|\psi|^6\cos6\phi$  & $g<0:G_1=C_s(\sigma_v);\ g>0:G_2=C_{s}(\sigma_d)$    &N&N&Y\\
 \cline{2-7}
 &$E_2$& $-w|\psi|^3\cos3\phi+g|\psi|^6\cos6\phi$  & Fig. \ref{phase36}: $G_1=C_{2v};\ G_2=C_{2}$   &Y&Y&N\\
 \hline
 \multirow{2}*{$D_{3h}$}& $E_1$& $g|\psi|^6\cos6\phi$  & $g<0:G_1=C_{s}(\sigma_v);\ g>0:G_2=C_{2}(C_{2}')$    &N&N&Y\\
 \cline{2-7}
 &$E_2$& $-w|\psi|^3\cos3\phi+g|\psi|^6\cos6\phi$  & Fig. \ref{phase36}: $G_1=C_{2v};\ G_2=C_s(\sigma_h)$   &Y&Y&N\\
 \hline
 \multicolumn{2}{|c||}{IO} &\multicolumn{5}{c|}{$ E_1E_2B_1,\ E_1E_2B_2  $}  \\
 \hline
\end{tabular}
\caption{Table for isomorphic groups $D_6$, $C_{6v}$ and $D_{3h}$. The notations are introduced in Section \ref{notation}.}
\label{table:D6}
\end{table*}

\subsection{Systems with $D_{6h}$ symmetry}

The behaviour of the even-parity irreps $E_{1g}$ and $E_{2g}$ in $D_{6h}$ are the same as those in $D_6$, with the only change that the symmetry $G_1$ and $G_2$ of the low temperature phase is replaced by the direct product of those in $D_6$ and group $C_i$.

For irrep $E_{1u}$ and $E_{2u}$, the free energy has the same form
\be
f=f_{iso}+g|\psi|^6\cos6\phi,
\ee
where we choose $\psi=|\psi|e^{i\phi}\sim x+iy$ for $E_{1u}$ and $\psi\sim (x^3-3xy^2)(x+iy)A_{1u}$ for $E_{2u}$. Emergent continuous symmetry exists in this system. There are two distinct low temperature phases for both irreps. If $g<0$, the low temperature phase has $\phi=\frac{n\pi}{3}$, and the symmetry is broken to subgroup $G_1=C_{2v}(C_2'')$ for for $E_{1u}$, and $G_1=D_2$ for $E_{2u}$. If $g>0$, the low temperature phase has $\phi=\frac{n\pi}{3}+\frac{\pi}{6}$, and the symmetry is broken to subgroup $G_2=C_{2v}(C_2')$ for $E_{1u}$ and $G_2=C_{2v}$ for $E_{2u}$.

\begin{table*}
\centering
\begin{tabular}{ |c||c|c|c|c|c|  }
 \hline
 \multicolumn{6}{|c|}{$D_{6h}$} \\
 \hline
  & $f_A$ &  Symmetry of low temperature phases & FOT & PTS &ECS   \\
 \hline
 $E_{1g}$& $g|\psi|^6\cos6\phi$  & $g<0:G_1=C_{2h}(C_{2}'');\ g>0:G_2=C_{2h}(C_{2}') $   &N&N&Y\\
 \hline
 $E_{1u}$& $g|\psi|^6\cos6\phi$  & $g<0:G_1=C_{2v}(C_{2}'');\ g>0:G_2=C_{2v}(C_{2}') $   &N&N&Y\\
 \hline
 $E_{2g}$& $-w|\psi|^3\cos3\phi+g|\psi|^6\cos(6\phi)$  & Fig. \ref{phase36}: $G_1=D_{2h}; \ G_2=C_{2h}(C_{2}) $   &Y&Y&N\\
 \hline
 $E_{2u}$& $g|\psi|^6\cos6\phi$  & $\lambda>0: G_1=D_2; \ \lambda<0:G_2=C_{2v}(C_{2}) $   &N&N&Y\\
 \hline
 IO &\multicolumn{5}{c|}{$\{ E_1E_2B_1\},\{ E_1E_2B_2 \},E_{1g}E_{1u}A_{2u},E_{1g}E_{1u}A_{1u},E_{2g}E_{2u}A_{1u},E_{2g}E_{2u}A_{2u},B_{1g}B_{1u}A_{1u},B_{2g}B_{2u}A_{1u},A_{2g}A_{2u}A_{1u}$}  \\
 \hline
\end{tabular}
\caption{Table for group $D_{6h}$. The notations are introduced in Section \ref{notation}.   }
\label{table_D6h}
\end{table*}

\subsection{Systems with $D_{4h}$ symmetry}

There are two 2D irreps $E_g$ and $E_u$, with order parameters that transform as the x and y components of pseudo vectors and polar vectors. Their free energy has the same form
\be
f=f_{iso}+\lambda(\psi_x^2-\psi_y^2)^2
\ee
The phase transition is second order. There are two low energy phases with different symmetries depending on the sign of $\lambda$. If $\lambda>0$, at low temperature the free energy is lowest when $\psi_x=\pm\psi_y$. In this phase, for order parameter in $E_g$, the symmetry reduces from $D_{4h}$ to $G_1=C_{2h}(C_2'')$, and for $E_u$ the symmetry reduces to $G_1=C_{2v}(C_2'')$. If $\lambda<0$, the low temperature phase has $\psi_x=0$ or $\psi_y=0$. In this phase, for $E_g$ the symmetry reduces from $D_{4h}$ to $G_2=C_{2h}(C_2')$, and for $E_u$ the symmetry reduces to $G_2=C_{2v}(C_2')$.

\begin{table*}
\centering
\begin{tabular}{ |c||c|c|c|c|c|  }
 \hline
 \multicolumn{6}{|c|}{$D_{4h}$} \\
 \hline
  & $f_A$ &  Symmetry of low temperature phases & FOT & PTS &ECS   \\
 \hline
 $E_g$& $\lambda(\psi_x^2-\psi_y^2)^2$  & $\lambda>0:G_1=C_{2h}(C_2'');\ \lambda<0:G_{2}=C_{2h}(C_{2}')$    &N&N&N\\
 \hline
 $E_u$& $\lambda(\psi_x^2-\psi_y^2)^2$  & $\lambda>0:G_1=C_{2v}(C_2'');\ \lambda<0:G_2=C_{2v}(C_{2}')$    &N&N&N\\
 \hline
 IO &\multicolumn{5}{c|}{$ \{A_2B_1B_2 \},E_gE_uA_{1u},E_gE_uA_{2u},E_gE_uB_{1u},E_gE_uB_{2u},B_{1g}B_{1u}A_{1u},B_{2g}B_{2u}A_{1u},A_{2g}A_{2u}A_{1u}  $}  \\
 \hline
\end{tabular}
\caption{ Table for group $D_{4h}$. The notations are introduced in Section \ref{notation}.}
\label{table_D4h}
\end{table*}

\subsection{The rest of the groups}

The examples given above about the high symmetry groups covers most of the physical phenomena that we focus on. We combine the results of the rest of the groups to a single table in table \ref{table:combined}. It can be seen that all the rotation groups $C_n$ and $C_n\times C_i$ have only one symmetry type for the low temperature phase, because the symmetry is too low to support multiple ways of symmetry breaking. These tables cover 24 of the 32 point groups. The groups that are not in these tables are $D_2$, $D_{2h}$, $C_2$, $C_{2v}$, $C_{2h}$, $C_s$, $C_i$ and $C_1$. These groups have only one-dimensional irreps, which implies they anisotropic term in free energy $f_A$ is zero and the entries for FOT, PTS and ECS are all negative.

\begin{table*}
\centering
\begin{tabular}{ |c|c||c|c|c|c|c|  }
 \hline
  & & $f_A$ &  Symmetry of low temperature phases & FOT & PTS &ECS   \\
 \hline
 \multirow{2}*{$C_6$}& $E_1$& $g|\psi|^6\cos(6\phi+\delta)$  & $G=C_1$    &N&N&Y\\
 \cline{2-7}
 &$E_2$& $-w|\psi|^3\cos(3\phi+\delta)+g|\psi|^6\cos(6\phi+\delta')$  & $G=C_2$    &Y&N&N\\
 \hline
 \multirow{2}*{$C_{3h}$}& $E_1$& $g|\psi|^6\cos(6\phi+\delta)$  & $G=C_1$    &N&N&Y\\
 \cline{2-7}
 &$E_2$& $-w|\psi|^3\cos(3\phi+\delta)+g|\psi|^6\cos(6\phi+\delta')$  & $G=C_s$    &Y&N&N\\
 \hline
 \multicolumn{2}{|c||}{IO} &\multicolumn{5}{c|}{$ BE_1E_2  $}  \\
 \hline
 \hline
 \multirow{4}*{$C_{6h}$}& $E_{1g}$& $g|\psi|^6\cos(6\phi+\delta)$  & $G=C_i $   &N&N&Y\\
 \cline{2-7}
 &$E_{1u}$& $g|\psi|^6\cos(6\phi+\delta)$  & $G=C_s $   &N&N&Y\\
 \cline{2-7}
 &$E_{2g}$& $-w|\psi|^3\cos(3\phi+\delta)+g|\psi|^6\cos(6\phi+\delta')$  & $G=C_{2h} $   &Y&N&N\\
 \cline{2-7}
 &$E_{2u}$& $g|\psi|^6\cos(6\phi+\delta)$  & $G=C_2$   &N&N&Y\\
 \hline
 \multicolumn{2}{|c||}{IO} &\multicolumn{5}{c|}{$ \{BE_1E_2\}, E_{1g}E_{1u}E_{2u},B_gB_uA_u $}  \\
 \hline
 \hline
 \multirow{1}*{$D_3$}& $E$& $-w|\psi|^3\cos3\phi+g|\psi|^6\cos6\phi$  & Fig. \ref{phase36}: $G_1=C_2; \ G_2=C_1 $    &Y&Y&N\\
 \hline
 \multirow{1}*{$C_{3v}$}& $E$& $-w|\psi|^3\cos3\phi+g|\psi|^6\cos6\phi$  & Fig. \ref{phase36}: $G_1=C_s; \ G_2=C_1 $    &Y&Y&N\\
 \hline
 \multicolumn{2}{|c||}{IO} &\multicolumn{5}{c|}{ None }  \\
 \hline
 \hline
 \multirow{2}*{$D_{3d}$}& $E_{g}$& $-w|\psi|^3\cos3\phi+g|\psi|^6\cos6\phi$  & Fig. \ref{phase36}: $=C_{2h}; \ G_2=C_i $    &Y&Y&N\\
 \cline{2-7}
 &$E_{u}$& $g|\psi|^6\cos6\phi$  & $ g<0:G_1=C_2;\ g>0:G_2=C_s $    &N&N&Y\\
 \hline
 \multicolumn{2}{|c||}{IO} &\multicolumn{5}{c|}{$ E_gE_uA_{1u},\ E_gE_uA_{2u}  $}  \\
 \hline
 \hline
 \multirow{1}*{$C_3$}& $E$& $-w|\psi|^3\cos(3\phi+\delta)+g|\psi|^6\cos(6\phi+\delta')$  & $G=C_1 $    &Y&N&N\\
 \hline
 \multicolumn{2}{|c||}{IO} &\multicolumn{5}{c|}{ None }  \\
 \hline
 \hline
 \multirow{2}*{$C_{3i}$}& $E_{g}$& $-w|\psi|^3\cos(3\phi+\delta)+g|\psi|^6\cos(6\phi+\delta')$  & $G=C_i $    &Y&N&N\\
 \cline{2-7}
 &$E_{u}$& $g|\psi|^6\cos(6\phi+\delta')$  &$ G=C_1 $    &N&N&Y\\
 \hline
 \multicolumn{2}{|c||}{IO} &\multicolumn{5}{c|}{None}  \\
 \hline
 \hline
 \multirow{1}*{$D_4$}& $E$& $\lambda(\psi_x^2-\psi_y^2)^2$  & $\lambda>0:G_1=C_2(C_2'');\ \lambda<0:G_2=C_2(C_{2}')$    &N&N&N\\
 \hline
 \multirow{1}*{$C_{4v}$}& $E$& $\lambda(\psi_x^2-\psi_y^2)^2$  & $\lambda>0:G_1=C_s(\sigma_d);\ \lambda<0:G_2=C_s(\sigma_v)$    &N&N&N\\
 \hline
 \multirow{1}*{$D_{2d}$}& $E$&  $\lambda(\psi_x^2-\psi_y^2)^2$  & $\lambda>0:G_1=C_s(\sigma_d);\ \lambda<0:G_2=C_2(C_{2}')$    &N&N&N\\
 \hline
 \multicolumn{2}{|c||}{IO} &\multicolumn{5}{c|}{$ A_2B_1B_2  $}  \\
 \hline
 \hline
 \multirow{1}*{$C_4$}& $E$& $\lambda|\psi|^4\cos(4\phi+\delta)$  & $G=C_1$    &N&N&N\\
 \hline
 \multirow{1}*{$S_4$}& $E$& $\lambda|\psi|^4\cos(4\phi+\delta)$  & $G=C_1$    &N&N&N\\
 \hline
 \multicolumn{2}{|c||}{IO} &\multicolumn{5}{c|}{None}  \\
 \hline
 \hline
 \multirow{2}*{$C_{4h}$} & $E_g$& $\lambda|\psi|^4\cos(4\phi+\delta)$  & $G=C_i$    &N&N&N\\
 \cline{2-7}
 &$E_u$& $\lambda|\psi|^4\cos(4\phi+\delta)$  & $G=C_1$    &N&N&N\\
 \hline
 \multicolumn{2}{|c||}{IO} &\multicolumn{5}{c|}{$ E_gE_uB_{u},\ B_gB_uA_u  $}  \\
 \hline
\end{tabular}
\caption{Table for systems with other point group symmetries. The notations are introduced in Section \ref{notation}. The isomorphic point groups are listed in the same block of the table. They share the same features in phase transition. Among these groups, the isomorphic ones are: $C_6$ and $C_{3h}$; $D_3$ and $C_{3v}$; $D_4$, $C_{4v}$ and $D_{2d}$; $C_4$ and $S_4$.  }
\label{table:combined}
\end{table*}

\section{Conclusion}

We present a comprehensive classification for phase transitions in the presence of point group symmetries. We analyzed the emergent phenomena in phase transition, such as the phase transition sequence, emergent symmetry and intertwined order. In particular, we established a mapping between symmetry breaking and the nature of phase transition, as in Fig. \ref{gpattern}. The symmetry before and after the phase transition uniquely determines the irrep of order parameter and the order of phase transition. For each order parameter that transforms under a given irrep, we provided the generic form of Ginzburg-Landau free energy compatible with symmetry and classified the possible symmetry-broken phases that the order parameter can lead to. This finding will be helpful to determine the type of phase transition from symmetry breaking in experiments.

\section{Acknowledgement}

H.L. acknowledges the fruitful discussion with K. Sun and the support of the National Science Foundation Grant No. EFRI-1741618.


\begin{thebibliography}{33}%
\makeatletter
\providecommand \@ifxundefined [1]{%
 \@ifx{#1\undefined}
}%
\providecommand \@ifnum [1]{%
 \ifnum #1\expandafter \@firstoftwo
 \else \expandafter \@secondoftwo
 \fi
}%
\providecommand \@ifx [1]{%
 \ifx #1\expandafter \@firstoftwo
 \else \expandafter \@secondoftwo
 \fi
}%
\providecommand \natexlab [1]{#1}%
\providecommand \enquote  [1]{``#1''}%
\providecommand \bibnamefont  [1]{#1}%
\providecommand \bibfnamefont [1]{#1}%
\providecommand \citenamefont [1]{#1}%
\providecommand \href@noop [0]{\@secondoftwo}%
\providecommand \href [0]{\begingroup \@sanitize@url \@href}%
\providecommand \@href[1]{\@@startlink{#1}\@@href}%
\providecommand \@@href[1]{\endgroup#1\@@endlink}%
\providecommand \@sanitize@url [0]{\catcode `\\12\catcode `\$12\catcode
  `\&12\catcode `\#12\catcode `\^12\catcode `\_12\catcode `\%12\relax}%
\providecommand \@@startlink[1]{}%
\providecommand \@@endlink[0]{}%
\providecommand \url  [0]{\begingroup\@sanitize@url \@url }%
\providecommand \@url [1]{\endgroup\@href {#1}{\urlprefix }}%
\providecommand \urlprefix  [0]{URL }%
\providecommand \Eprint [0]{\href }%
\providecommand \doibase [0]{http://dx.doi.org/}%
\providecommand \selectlanguage [0]{\@gobble}%
\providecommand \bibinfo  [0]{\@secondoftwo}%
\providecommand \bibfield  [0]{\@secondoftwo}%
\providecommand \translation [1]{[#1]}%
\providecommand \BibitemOpen [0]{}%
\providecommand \bibitemStop [0]{}%
\providecommand \bibitemNoStop [0]{.\EOS\space}%
\providecommand \EOS [0]{\spacefactor3000\relax}%
\providecommand \BibitemShut  [1]{\csname bibitem#1\endcsname}%
\let\auto@bib@innerbib\@empty
\bibitem [{\citenamefont {Landau}(1937)}]{Landau1937}%
  \BibitemOpen
  \bibfield  {author} {\bibinfo {author} {\bibfnamefont {L.~D.}\ \bibnamefont
  {Landau}},\ }\bibfield  {title} {\enquote {\bibinfo {title} {{On the theory
  of phase transitions}},}\ }\href@noop {} {\bibfield  {journal} {\bibinfo
  {journal} {Zh. Eksp. Teor. Fiz.}\ }\textbf {\bibinfo {volume} {7}},\ \bibinfo
  {pages} {19--32} (\bibinfo {year} {1937})}\BibitemShut {NoStop}%
\bibitem [{\citenamefont {De~Gennes}\ and\ \citenamefont
  {Prost}(1993)}]{nematicbook1993}%
  \BibitemOpen
  \bibfield  {author} {\bibinfo {author} {\bibfnamefont {Pierre-Gilles}\
  \bibnamefont {De~Gennes}}\ and\ \bibinfo {author} {\bibfnamefont {Jacques}\
  \bibnamefont {Prost}},\ }\href@noop {} {\emph {\bibinfo {title} {The physics
  of liquid crystals}}},\ \bibinfo {number} {83}\ (\bibinfo  {publisher}
  {Oxford university press},\ \bibinfo {year} {1993})\BibitemShut {NoStop}%
\bibitem [{\citenamefont {Chaikin}\ \emph {et~al.}(1995)\citenamefont
  {Chaikin}, \citenamefont {Lubensky},\ and\ \citenamefont
  {Witten}}]{chaikin1995principles}%
  \BibitemOpen
  \bibfield  {author} {\bibinfo {author} {\bibfnamefont {Paul~M}\ \bibnamefont
  {Chaikin}}, \bibinfo {author} {\bibfnamefont {Tom~C}\ \bibnamefont
  {Lubensky}}, \ and\ \bibinfo {author} {\bibfnamefont {Thomas~A}\ \bibnamefont
  {Witten}},\ }\href@noop {} {\emph {\bibinfo {title} {Principles of condensed
  matter physics}}},\ Vol.~\bibinfo {volume} {10}\ (\bibinfo  {publisher}
  {Cambridge university press Cambridge},\ \bibinfo {year} {1995})\BibitemShut
  {NoStop}%
\bibitem [{\citenamefont {Kivelson}\ \emph {et~al.}(1998)\citenamefont
  {Kivelson}, \citenamefont {Fradkin},\ and\ \citenamefont
  {Emery}}]{Kivelson1998}%
  \BibitemOpen
  \bibfield  {author} {\bibinfo {author} {\bibfnamefont {S.~A.}\ \bibnamefont
  {Kivelson}}, \bibinfo {author} {\bibfnamefont {E.}~\bibnamefont {Fradkin}}, \
  and\ \bibinfo {author} {\bibfnamefont {V.~J.}\ \bibnamefont {Emery}},\
  }\bibfield  {title} {\enquote {\bibinfo {title} {Electronic liquid-crystal
  phases of a doped mott insulator},}\ }\href {\doibase 10.1038/31177}
  {\bibfield  {journal} {\bibinfo  {journal} {Nature}\ }\textbf {\bibinfo
  {volume} {393}},\ \bibinfo {pages} {550--553} (\bibinfo {year}
  {1998})}\BibitemShut {NoStop}%
\bibitem [{\citenamefont {Fradkin}\ and\ \citenamefont
  {Kivelson}(1999)}]{Fradkin1999}%
  \BibitemOpen
  \bibfield  {author} {\bibinfo {author} {\bibfnamefont {Eduardo}\ \bibnamefont
  {Fradkin}}\ and\ \bibinfo {author} {\bibfnamefont {Steven~A.}\ \bibnamefont
  {Kivelson}},\ }\bibfield  {title} {\enquote {\bibinfo {title} {Liquid-crystal
  phases of quantum hall systems},}\ }\href {\doibase 10.1103/PhysRevB.59.8065}
  {\bibfield  {journal} {\bibinfo  {journal} {Phys. Rev. B}\ }\textbf {\bibinfo
  {volume} {59}},\ \bibinfo {pages} {8065--8072} (\bibinfo {year}
  {1999})}\BibitemShut {NoStop}%
\bibitem [{\citenamefont {Oganesyan}\ \emph {et~al.}(2001)\citenamefont
  {Oganesyan}, \citenamefont {Kivelson},\ and\ \citenamefont
  {Fradkin}}]{Oganesyan2001}%
  \BibitemOpen
  \bibfield  {author} {\bibinfo {author} {\bibfnamefont {Vadim}\ \bibnamefont
  {Oganesyan}}, \bibinfo {author} {\bibfnamefont {Steven~A.}\ \bibnamefont
  {Kivelson}}, \ and\ \bibinfo {author} {\bibfnamefont {Eduardo}\ \bibnamefont
  {Fradkin}},\ }\bibfield  {title} {\enquote {\bibinfo {title} {Quantum theory
  of a nematic fermi fluid},}\ }\href {\doibase 10.1103/PhysRevB.64.195109}
  {\bibfield  {journal} {\bibinfo  {journal} {Phys. Rev. B}\ }\textbf {\bibinfo
  {volume} {64}},\ \bibinfo {pages} {195109} (\bibinfo {year}
  {2001})}\BibitemShut {NoStop}%
\bibitem [{\citenamefont {Kimura}\ \emph {et~al.}(2003)\citenamefont {Kimura},
  \citenamefont {Goto}, \citenamefont {Shintani}, \citenamefont {Ishizaka},
  \citenamefont {Arima},\ and\ \citenamefont {Tokura}}]{Kimura2003}%
  \BibitemOpen
  \bibfield  {author} {\bibinfo {author} {\bibfnamefont {T.}~\bibnamefont
  {Kimura}}, \bibinfo {author} {\bibfnamefont {T.}~\bibnamefont {Goto}},
  \bibinfo {author} {\bibfnamefont {H.}~\bibnamefont {Shintani}}, \bibinfo
  {author} {\bibfnamefont {K.}~\bibnamefont {Ishizaka}}, \bibinfo {author}
  {\bibfnamefont {T.}~\bibnamefont {Arima}}, \ and\ \bibinfo {author}
  {\bibfnamefont {Y.}~\bibnamefont {Tokura}},\ }\bibfield  {title} {\enquote
  {\bibinfo {title} {Magnetic control of ferroelectric polarization},}\ }\href
  {\doibase 10.1038/nature02018} {\bibfield  {journal} {\bibinfo  {journal}
  {Nature}\ }\textbf {\bibinfo {volume} {426}},\ \bibinfo {pages} {55--58}
  (\bibinfo {year} {2003})}\BibitemShut {NoStop}%
\bibitem [{\citenamefont {Cheong}\ and\ \citenamefont
  {Mostovoy}(2007)}]{Cheong2007}%
  \BibitemOpen
  \bibfield  {author} {\bibinfo {author} {\bibfnamefont {Sang-Wook}\
  \bibnamefont {Cheong}}\ and\ \bibinfo {author} {\bibfnamefont {Maxim}\
  \bibnamefont {Mostovoy}},\ }\bibfield  {title} {\enquote {\bibinfo {title}
  {Multiferroics: a magnetic twist for ferroelectricity},}\ }\href {\doibase
  10.1038/nmat1804} {\bibfield  {journal} {\bibinfo  {journal} {Nature
  Materials}\ }\textbf {\bibinfo {volume} {6}},\ \bibinfo {pages} {13--20}
  (\bibinfo {year} {2007})}\BibitemShut {NoStop}%
\bibitem [{\citenamefont {Cheong}\ \emph {et~al.}(2018)\citenamefont {Cheong},
  \citenamefont {Talbayev}, \citenamefont {Kiryukhin},\ and\ \citenamefont
  {Saxena}}]{Cheong2018}%
  \BibitemOpen
  \bibfield  {author} {\bibinfo {author} {\bibfnamefont {Sang-Wook}\
  \bibnamefont {Cheong}}, \bibinfo {author} {\bibfnamefont {Diyar}\
  \bibnamefont {Talbayev}}, \bibinfo {author} {\bibfnamefont {Valery}\
  \bibnamefont {Kiryukhin}}, \ and\ \bibinfo {author} {\bibfnamefont {Avadh}\
  \bibnamefont {Saxena}},\ }\bibfield  {title} {\enquote {\bibinfo {title}
  {Broken symmetries, non-reciprocity, and multiferroicity},}\ }\href {\doibase
  10.1038/s41535-018-0092-5} {\bibfield  {journal} {\bibinfo  {journal} {npj
  Quantum Materials}\ }\textbf {\bibinfo {volume} {3}},\ \bibinfo {pages} {19}
  (\bibinfo {year} {2018})}\BibitemShut {NoStop}%
\bibitem [{\citenamefont {Van~Aken}\ \emph {et~al.}(2007)\citenamefont
  {Van~Aken}, \citenamefont {Rivera}, \citenamefont {Schmid},\ and\
  \citenamefont {Fiebig}}]{VanAken2007}%
  \BibitemOpen
  \bibfield  {author} {\bibinfo {author} {\bibfnamefont {Bas~B.}\ \bibnamefont
  {Van~Aken}}, \bibinfo {author} {\bibfnamefont {Jean-Pierre}\ \bibnamefont
  {Rivera}}, \bibinfo {author} {\bibfnamefont {Hans}\ \bibnamefont {Schmid}}, \
  and\ \bibinfo {author} {\bibfnamefont {Manfred}\ \bibnamefont {Fiebig}},\
  }\bibfield  {title} {\enquote {\bibinfo {title} {Observation of ferrotoroidic
  domains},}\ }\href {\doibase 10.1038/nature06139} {\bibfield  {journal}
  {\bibinfo  {journal} {Nature}\ }\textbf {\bibinfo {volume} {449}},\ \bibinfo
  {pages} {702--705} (\bibinfo {year} {2007})}\BibitemShut {NoStop}%
\bibitem [{\citenamefont {Spaldin}\ \emph {et~al.}(2008)\citenamefont
  {Spaldin}, \citenamefont {Fiebig},\ and\ \citenamefont
  {Mostovoy}}]{Spaldin_2008}%
  \BibitemOpen
  \bibfield  {author} {\bibinfo {author} {\bibfnamefont {Nicola~A}\
  \bibnamefont {Spaldin}}, \bibinfo {author} {\bibfnamefont {Manfred}\
  \bibnamefont {Fiebig}}, \ and\ \bibinfo {author} {\bibfnamefont {Maxim}\
  \bibnamefont {Mostovoy}},\ }\bibfield  {title} {\enquote {\bibinfo {title}
  {The toroidal moment in condensed-matter physics and its relation to the
  magnetoelectric effect},}\ }\href {\doibase 10.1088/0953-8984/20/43/434203}
  {\bibfield  {journal} {\bibinfo  {journal} {Journal of Physics: Condensed
  Matter}\ }\textbf {\bibinfo {volume} {20}},\ \bibinfo {pages} {434203}
  (\bibinfo {year} {2008})}\BibitemShut {NoStop}%
\bibitem [{\citenamefont {Zimmermann}\ \emph {et~al.}(2014)\citenamefont
  {Zimmermann}, \citenamefont {Meier},\ and\ \citenamefont
  {Fiebig}}]{Zimmermann2014}%
  \BibitemOpen
  \bibfield  {author} {\bibinfo {author} {\bibfnamefont {Anne~S.}\ \bibnamefont
  {Zimmermann}}, \bibinfo {author} {\bibfnamefont {Dennis}\ \bibnamefont
  {Meier}}, \ and\ \bibinfo {author} {\bibfnamefont {Manfred}\ \bibnamefont
  {Fiebig}},\ }\bibfield  {title} {\enquote {\bibinfo {title} {Ferroic nature
  of magnetic toroidal order},}\ }\href {\doibase 10.1038/ncomms5796}
  {\bibfield  {journal} {\bibinfo  {journal} {Nature Communications}\ }\textbf
  {\bibinfo {volume} {5}},\ \bibinfo {pages} {4796} (\bibinfo {year}
  {2014})}\BibitemShut {NoStop}%
\bibitem [{\citenamefont {Gopalan}\ and\ \citenamefont
  {Litvin}(2011)}]{Gopalan2011}%
  \BibitemOpen
  \bibfield  {author} {\bibinfo {author} {\bibfnamefont {Venkatraman}\
  \bibnamefont {Gopalan}}\ and\ \bibinfo {author} {\bibfnamefont {Daniel~B.}\
  \bibnamefont {Litvin}},\ }\bibfield  {title} {\enquote {\bibinfo {title}
  {Rotation-reversal symmetries in crystals and handed structures},}\ }\href
  {\doibase 10.1038/nmat2987} {\bibfield  {journal} {\bibinfo  {journal}
  {Nature Materials}\ }\textbf {\bibinfo {volume} {10}},\ \bibinfo {pages}
  {376--381} (\bibinfo {year} {2011})}\BibitemShut {NoStop}%
\bibitem [{\citenamefont {Johnson}\ \emph {et~al.}(2012)\citenamefont
  {Johnson}, \citenamefont {Chapon}, \citenamefont {Khalyavin}, \citenamefont
  {Manuel}, \citenamefont {Radaelli},\ and\ \citenamefont
  {Martin}}]{Johnson2012}%
  \BibitemOpen
  \bibfield  {author} {\bibinfo {author} {\bibfnamefont {R.~D.}\ \bibnamefont
  {Johnson}}, \bibinfo {author} {\bibfnamefont {L.~C.}\ \bibnamefont {Chapon}},
  \bibinfo {author} {\bibfnamefont {D.~D.}\ \bibnamefont {Khalyavin}}, \bibinfo
  {author} {\bibfnamefont {P.}~\bibnamefont {Manuel}}, \bibinfo {author}
  {\bibfnamefont {P.~G.}\ \bibnamefont {Radaelli}}, \ and\ \bibinfo {author}
  {\bibfnamefont {C.}~\bibnamefont {Martin}},\ }\bibfield  {title} {\enquote
  {\bibinfo {title} {Giant improper ferroelectricity in the ferroaxial magnet
  ${\mathrm{camn}}_{7}{\mathbf{o}}_{12}$},}\ }\href {\doibase
  10.1103/PhysRevLett.108.067201} {\bibfield  {journal} {\bibinfo  {journal}
  {Phys. Rev. Lett.}\ }\textbf {\bibinfo {volume} {108}},\ \bibinfo {pages}
  {067201} (\bibinfo {year} {2012})}\BibitemShut {NoStop}%
\bibitem [{\citenamefont {Terada}\ \emph {et~al.}(2015)\citenamefont {Terada},
  \citenamefont {Khalyavin}, \citenamefont {Manuel}, \citenamefont {Yi},
  \citenamefont {Suzuki}, \citenamefont {Tsujii}, \citenamefont {Imanaka},\
  and\ \citenamefont {Belik}}]{Terada2015}%
  \BibitemOpen
  \bibfield  {author} {\bibinfo {author} {\bibfnamefont {Noriki}\ \bibnamefont
  {Terada}}, \bibinfo {author} {\bibfnamefont {Dmitry~D.}\ \bibnamefont
  {Khalyavin}}, \bibinfo {author} {\bibfnamefont {Pascal}\ \bibnamefont
  {Manuel}}, \bibinfo {author} {\bibfnamefont {Wei}\ \bibnamefont {Yi}},
  \bibinfo {author} {\bibfnamefont {Hiroyuki~S.}\ \bibnamefont {Suzuki}},
  \bibinfo {author} {\bibfnamefont {Naohito}\ \bibnamefont {Tsujii}}, \bibinfo
  {author} {\bibfnamefont {Yasutaka}\ \bibnamefont {Imanaka}}, \ and\ \bibinfo
  {author} {\bibfnamefont {Alexei~A.}\ \bibnamefont {Belik}},\ }\bibfield
  {title} {\enquote {\bibinfo {title} {Ferroelectricity induced by ferriaxial
  crystal rotation and spin helicity in a $b$-site-ordered double-perovskite
  multiferroic ${\text{in}}_{2}{\text{nimno}}_{6}$},}\ }\href {\doibase
  10.1103/PhysRevB.91.104413} {\bibfield  {journal} {\bibinfo  {journal} {Phys.
  Rev. B}\ }\textbf {\bibinfo {volume} {91}},\ \bibinfo {pages} {104413}
  (\bibinfo {year} {2015})}\BibitemShut {NoStop}%
\bibitem [{\citenamefont {Jin}\ \emph {et~al.}(2020)\citenamefont {Jin},
  \citenamefont {Drueke}, \citenamefont {Li}, \citenamefont {Admasu},
  \citenamefont {Owen}, \citenamefont {Day}, \citenamefont {Sun}, \citenamefont
  {Cheong},\ and\ \citenamefont {Zhao}}]{Jin2020}%
  \BibitemOpen
  \bibfield  {author} {\bibinfo {author} {\bibfnamefont {Wencan}\ \bibnamefont
  {Jin}}, \bibinfo {author} {\bibfnamefont {Elizabeth}\ \bibnamefont {Drueke}},
  \bibinfo {author} {\bibfnamefont {Siwen}\ \bibnamefont {Li}}, \bibinfo
  {author} {\bibfnamefont {Alemayehu}\ \bibnamefont {Admasu}}, \bibinfo
  {author} {\bibfnamefont {Rachel}\ \bibnamefont {Owen}}, \bibinfo {author}
  {\bibfnamefont {Matthew}\ \bibnamefont {Day}}, \bibinfo {author}
  {\bibfnamefont {Kai}\ \bibnamefont {Sun}}, \bibinfo {author} {\bibfnamefont
  {Sang-Wook}\ \bibnamefont {Cheong}}, \ and\ \bibinfo {author} {\bibfnamefont
  {Liuyan}\ \bibnamefont {Zhao}},\ }\bibfield  {title} {\enquote {\bibinfo
  {title} {Observation of a ferro-rotational order coupled with second-order
  nonlinear optical fields},}\ }\href {\doibase 10.1038/s41567-019-0695-1}
  {\bibfield  {journal} {\bibinfo  {journal} {Nature Physics}\ }\textbf
  {\bibinfo {volume} {16}},\ \bibinfo {pages} {42--46} (\bibinfo {year}
  {2020})}\BibitemShut {NoStop}%
\bibitem [{\citenamefont {Hahn}\ \emph {et~al.}(1983)\citenamefont {Hahn},
  \citenamefont {Shmueli},\ and\ \citenamefont
  {Arthur}}]{hahn1983international}%
  \BibitemOpen
  \bibfield  {author} {\bibinfo {author} {\bibfnamefont {Theo}\ \bibnamefont
  {Hahn}}, \bibinfo {author} {\bibfnamefont {Uri}\ \bibnamefont {Shmueli}}, \
  and\ \bibinfo {author} {\bibfnamefont {JC~Wilson}\ \bibnamefont {Arthur}},\
  }\href@noop {} {\emph {\bibinfo {title} {International tables for
  crystallography}}},\ Vol.~\bibinfo {volume} {1}\ (\bibinfo  {publisher}
  {Reidel Dordrecht},\ \bibinfo {year} {1983})\BibitemShut {NoStop}%
\bibitem [{\citenamefont {Castellan}\ \emph {et~al.}(2002)\citenamefont
  {Castellan}, \citenamefont {Gaulin}, \citenamefont {van Duijn}, \citenamefont
  {Lewis}, \citenamefont {Lumsden}, \citenamefont {Jin}, \citenamefont {He},
  \citenamefont {Nagler},\ and\ \citenamefont {Mandrus}}]{Castellan2002}%
  \BibitemOpen
  \bibfield  {author} {\bibinfo {author} {\bibfnamefont {J.~P.}\ \bibnamefont
  {Castellan}}, \bibinfo {author} {\bibfnamefont {B.~D.}\ \bibnamefont
  {Gaulin}}, \bibinfo {author} {\bibfnamefont {J.}~\bibnamefont {van Duijn}},
  \bibinfo {author} {\bibfnamefont {M.~J.}\ \bibnamefont {Lewis}}, \bibinfo
  {author} {\bibfnamefont {M.~D.}\ \bibnamefont {Lumsden}}, \bibinfo {author}
  {\bibfnamefont {R.}~\bibnamefont {Jin}}, \bibinfo {author} {\bibfnamefont
  {J.}~\bibnamefont {He}}, \bibinfo {author} {\bibfnamefont {S.~E.}\
  \bibnamefont {Nagler}}, \ and\ \bibinfo {author} {\bibfnamefont
  {D.}~\bibnamefont {Mandrus}},\ }\bibfield  {title} {\enquote {\bibinfo
  {title} {Structural ordering and symmetry breaking in
  ${\mathrm{cd}}_{2}{\mathrm{re}}_{2}{\mathrm{o}}_{7}$},}\ }\href {\doibase
  10.1103/PhysRevB.66.134528} {\bibfield  {journal} {\bibinfo  {journal} {Phys.
  Rev. B}\ }\textbf {\bibinfo {volume} {66}},\ \bibinfo {pages} {134528}
  (\bibinfo {year} {2002})}\BibitemShut {NoStop}%
\bibitem [{\citenamefont {{de Vries}}(1986)}]{DEVRIES1986193}%
  \BibitemOpen
  \bibfield  {author} {\bibinfo {author} {\bibfnamefont {Adriaan}\ \bibnamefont
  {{de Vries}}},\ }\bibfield  {title} {\enquote {\bibinfo {title} {X-ray
  diffraction studies of the structure of the skewed cybotactic nematic phase:
  A review of the literature},}\ }\href {\doibase
  https://doi.org/10.1016/0167-7322(86)80001-0} {\bibfield  {journal} {\bibinfo
   {journal} {Journal of Molecular Liquids}\ }\textbf {\bibinfo {volume}
  {31}},\ \bibinfo {pages} {193--202} (\bibinfo {year} {1986})}\BibitemShut
  {NoStop}%
\bibitem [{\citenamefont {Harter}\ \emph {et~al.}(2017)\citenamefont {Harter},
  \citenamefont {Zhao}, \citenamefont {Yan}, \citenamefont {Mandrus},\ and\
  \citenamefont {Hsieh}}]{Harter295}%
  \BibitemOpen
  \bibfield  {author} {\bibinfo {author} {\bibfnamefont {J.~W.}\ \bibnamefont
  {Harter}}, \bibinfo {author} {\bibfnamefont {Z.~Y.}\ \bibnamefont {Zhao}},
  \bibinfo {author} {\bibfnamefont {J.-Q.}\ \bibnamefont {Yan}}, \bibinfo
  {author} {\bibfnamefont {D.~G.}\ \bibnamefont {Mandrus}}, \ and\ \bibinfo
  {author} {\bibfnamefont {D.}~\bibnamefont {Hsieh}},\ }\bibfield  {title}
  {\enquote {\bibinfo {title} {A parity-breaking electronic nematic phase
  transition in the spin-orbit coupled metal cd2re2o7},}\ }\href {\doibase
  10.1126/science.aad1188} {\bibfield  {journal} {\bibinfo  {journal}
  {Science}\ }\textbf {\bibinfo {volume} {356}},\ \bibinfo {pages} {295--299}
  (\bibinfo {year} {2017})}\BibitemShut {NoStop}%
\bibitem [{\citenamefont {Ok}\ \emph {et~al.}(2006)\citenamefont {Ok},
  \citenamefont {Chi},\ and\ \citenamefont {Halasyamani}}]{B511119F}%
  \BibitemOpen
  \bibfield  {author} {\bibinfo {author} {\bibfnamefont {Kang~Min}\
  \bibnamefont {Ok}}, \bibinfo {author} {\bibfnamefont {Eun~Ok}\ \bibnamefont
  {Chi}}, \ and\ \bibinfo {author} {\bibfnamefont {P.~Shiv}\ \bibnamefont
  {Halasyamani}},\ }\bibfield  {title} {\enquote {\bibinfo {title} {Bulk
  characterization methods for non-centrosymmetric materials: second-harmonic
  generation{,} piezoelectricity{,} pyroelectricity{,} and ferroelectricity},}\
  }\href {\doibase 10.1039/B511119F} {\bibfield  {journal} {\bibinfo  {journal}
  {Chem. Soc. Rev.}\ }\textbf {\bibinfo {volume} {35}},\ \bibinfo {pages}
  {710--717} (\bibinfo {year} {2006})}\BibitemShut {NoStop}%
\bibitem [{\citenamefont {Matlack}\ \emph {et~al.}(2014)\citenamefont
  {Matlack}, \citenamefont {Kim}, \citenamefont {Jacobs},\ and\ \citenamefont
  {Qu}}]{Matlack2014}%
  \BibitemOpen
  \bibfield  {author} {\bibinfo {author} {\bibfnamefont {K.~H.}\ \bibnamefont
  {Matlack}}, \bibinfo {author} {\bibfnamefont {J.-Y.}\ \bibnamefont {Kim}},
  \bibinfo {author} {\bibfnamefont {L.~J.}\ \bibnamefont {Jacobs}}, \ and\
  \bibinfo {author} {\bibfnamefont {J.}~\bibnamefont {Qu}},\ }\bibfield
  {title} {\enquote {\bibinfo {title} {Review of second harmonic generation
  measurement techniques for material state determination in metals},}\ }\href
  {\doibase 10.1007/s10921-014-0273-5} {\bibfield  {journal} {\bibinfo
  {journal} {Journal of Nondestructive Evaluation}\ }\textbf {\bibinfo {volume}
  {34}},\ \bibinfo {pages} {273} (\bibinfo {year} {2014})}\BibitemShut
  {NoStop}%
\bibitem [{\citenamefont {Fradkin}\ \emph {et~al.}(2015)\citenamefont
  {Fradkin}, \citenamefont {Kivelson},\ and\ \citenamefont
  {Tranquada}}]{Fradkin2015}%
  \BibitemOpen
  \bibfield  {author} {\bibinfo {author} {\bibfnamefont {Eduardo}\ \bibnamefont
  {Fradkin}}, \bibinfo {author} {\bibfnamefont {Steven~A.}\ \bibnamefont
  {Kivelson}}, \ and\ \bibinfo {author} {\bibfnamefont {John~M.}\ \bibnamefont
  {Tranquada}},\ }\bibfield  {title} {\enquote {\bibinfo {title} {Colloquium:
  Theory of intertwined orders in high temperature superconductors},}\ }\href
  {\doibase 10.1103/RevModPhys.87.457} {\bibfield  {journal} {\bibinfo
  {journal} {Rev. Mod. Phys.}\ }\textbf {\bibinfo {volume} {87}},\ \bibinfo
  {pages} {457--482} (\bibinfo {year} {2015})}\BibitemShut {NoStop}%
\bibitem [{\citenamefont {Fernandes}\ \emph {et~al.}(2019)\citenamefont
  {Fernandes}, \citenamefont {Orth},\ and\ \citenamefont
  {Schmalian}}]{Fernandes2019}%
  \BibitemOpen
  \bibfield  {author} {\bibinfo {author} {\bibfnamefont {Rafael~M.}\
  \bibnamefont {Fernandes}}, \bibinfo {author} {\bibfnamefont {Peter~P.}\
  \bibnamefont {Orth}}, \ and\ \bibinfo {author} {\bibfnamefont {Jörg}\
  \bibnamefont {Schmalian}},\ }\bibfield  {title} {\enquote {\bibinfo {title}
  {Intertwined vestigial order in quantum materials: Nematicity and beyond},}\
  }\href {\doibase 10.1146/annurev-conmatphys-031218-013200} {\bibfield
  {journal} {\bibinfo  {journal} {Annual Review of Condensed Matter Physics}\
  }\textbf {\bibinfo {volume} {10}},\ \bibinfo {pages} {133--154} (\bibinfo
  {year} {2019})}\BibitemShut {NoStop}%
\bibitem [{\citenamefont {Blankschtein}\ \emph {et~al.}(1984)\citenamefont
  {Blankschtein}, \citenamefont {Ma}, \citenamefont {Berker}, \citenamefont
  {Grest},\ and\ \citenamefont {Soukoulis}}]{Blankschtein1984}%
  \BibitemOpen
  \bibfield  {author} {\bibinfo {author} {\bibfnamefont {Daniel}\ \bibnamefont
  {Blankschtein}}, \bibinfo {author} {\bibfnamefont {M.}~\bibnamefont {Ma}},
  \bibinfo {author} {\bibfnamefont {A.~Nihat}\ \bibnamefont {Berker}}, \bibinfo
  {author} {\bibfnamefont {Gary~S.}\ \bibnamefont {Grest}}, \ and\ \bibinfo
  {author} {\bibfnamefont {C.~M.}\ \bibnamefont {Soukoulis}},\ }\bibfield
  {title} {\enquote {\bibinfo {title} {Orderings of a stacked frustrated
  triangular system in three dimensions},}\ }\href {\doibase
  10.1103/PhysRevB.29.5250} {\bibfield  {journal} {\bibinfo  {journal} {Phys.
  Rev. B}\ }\textbf {\bibinfo {volume} {29}},\ \bibinfo {pages} {5250--5252}
  (\bibinfo {year} {1984})}\BibitemShut {NoStop}%
\bibitem [{\citenamefont {Isakov}\ and\ \citenamefont
  {Moessner}(2003)}]{Isakov2003}%
  \BibitemOpen
  \bibfield  {author} {\bibinfo {author} {\bibfnamefont {S.~V.}\ \bibnamefont
  {Isakov}}\ and\ \bibinfo {author} {\bibfnamefont {R.}~\bibnamefont
  {Moessner}},\ }\bibfield  {title} {\enquote {\bibinfo {title} {Interplay of
  quantum and thermal fluctuations in a frustrated magnet},}\ }\href {\doibase
  10.1103/PhysRevB.68.104409} {\bibfield  {journal} {\bibinfo  {journal} {Phys.
  Rev. B}\ }\textbf {\bibinfo {volume} {68}},\ \bibinfo {pages} {104409}
  (\bibinfo {year} {2003})}\BibitemShut {NoStop}%
\bibitem [{\citenamefont {Gazit}\ \emph {et~al.}(2018)\citenamefont {Gazit},
  \citenamefont {Assaad}, \citenamefont {Sachdev}, \citenamefont {Vishwanath},\
  and\ \citenamefont {Wang}}]{Gazit2018}%
  \BibitemOpen
  \bibfield  {author} {\bibinfo {author} {\bibfnamefont {Snir}\ \bibnamefont
  {Gazit}}, \bibinfo {author} {\bibfnamefont {Fakher~F.}\ \bibnamefont
  {Assaad}}, \bibinfo {author} {\bibfnamefont {Subir}\ \bibnamefont {Sachdev}},
  \bibinfo {author} {\bibfnamefont {Ashvin}\ \bibnamefont {Vishwanath}}, \ and\
  \bibinfo {author} {\bibfnamefont {Chong}\ \bibnamefont {Wang}},\ }\bibfield
  {title} {\enquote {\bibinfo {title} {Confinement transition of z2 gauge
  theories coupled to massless fermions: Emergent quantum chromodynamics and
  so(5) symmetry},}\ }\href {\doibase 10.1073/pnas.1806338115} {\bibfield
  {journal} {\bibinfo  {journal} {Proceedings of the National Academy of
  Sciences}\ }\textbf {\bibinfo {volume} {115}},\ \bibinfo {pages}
  {E6987--E6995} (\bibinfo {year} {2018})}\BibitemShut {NoStop}%
\bibitem [{\citenamefont {Hlinka}\ \emph {et~al.}(2016)\citenamefont {Hlinka},
  \citenamefont {Privratska}, \citenamefont {Ondrejkovic},\ and\ \citenamefont
  {Janovec}}]{Hlinka2016}%
  \BibitemOpen
  \bibfield  {author} {\bibinfo {author} {\bibfnamefont {J.}~\bibnamefont
  {Hlinka}}, \bibinfo {author} {\bibfnamefont {J.}~\bibnamefont {Privratska}},
  \bibinfo {author} {\bibfnamefont {P.}~\bibnamefont {Ondrejkovic}}, \ and\
  \bibinfo {author} {\bibfnamefont {V.}~\bibnamefont {Janovec}},\ }\bibfield
  {title} {\enquote {\bibinfo {title} {Symmetry guide to ferroaxial
  transitions},}\ }\href {\doibase 10.1103/PhysRevLett.116.177602} {\bibfield
  {journal} {\bibinfo  {journal} {Phys. Rev. Lett.}\ }\textbf {\bibinfo
  {volume} {116}},\ \bibinfo {pages} {177602} (\bibinfo {year}
  {2016})}\BibitemShut {NoStop}%
\bibitem [{\citenamefont {Watanabe}\ and\ \citenamefont
  {Yanase}(2018)}]{Watanabe2018}%
  \BibitemOpen
  \bibfield  {author} {\bibinfo {author} {\bibfnamefont {Hikaru}\ \bibnamefont
  {Watanabe}}\ and\ \bibinfo {author} {\bibfnamefont {Youichi}\ \bibnamefont
  {Yanase}},\ }\bibfield  {title} {\enquote {\bibinfo {title}
  {Group-theoretical classification of multipole order: Emergent responses and
  candidate materials},}\ }\href {\doibase 10.1103/PhysRevB.98.245129}
  {\bibfield  {journal} {\bibinfo  {journal} {Phys. Rev. B}\ }\textbf {\bibinfo
  {volume} {98}},\ \bibinfo {pages} {245129} (\bibinfo {year}
  {2018})}\BibitemShut {NoStop}%
\bibitem [{\citenamefont {Erb}\ and\ \citenamefont {Hlinka}(2018)}]{Erb2018}%
  \BibitemOpen
  \bibfield  {author} {\bibinfo {author} {\bibfnamefont {K.~C.}\ \bibnamefont
  {Erb}}\ and\ \bibinfo {author} {\bibfnamefont {J.}~\bibnamefont {Hlinka}},\
  }\bibfield  {title} {\enquote {\bibinfo {title} {Symmetry guide to chiroaxial
  transitions},}\ }\href {\doibase 10.1080/01411594.2018.1498498} {\bibfield
  {journal} {\bibinfo  {journal} {Phase Transitions}\ }\textbf {\bibinfo
  {volume} {91}},\ \bibinfo {pages} {953--958} (\bibinfo {year}
  {2018})}\BibitemShut {NoStop}%
\bibitem [{\citenamefont {Norman}(2020)}]{Norman2020}%
  \BibitemOpen
  \bibfield  {author} {\bibinfo {author} {\bibfnamefont {M.~R.}\ \bibnamefont
  {Norman}},\ }\bibfield  {title} {\enquote {\bibinfo {title} {Crystal
  structure of the inversion-breaking metal
  ${\mathrm{cd}}_{2}{\mathrm{re}}_{2}{\mathrm{o}}_{7}$},}\ }\href {\doibase
  10.1103/PhysRevB.101.045117} {\bibfield  {journal} {\bibinfo  {journal}
  {Phys. Rev. B}\ }\textbf {\bibinfo {volume} {101}},\ \bibinfo {pages}
  {045117} (\bibinfo {year} {2020})}\BibitemShut {NoStop}%
\bibitem [{\citenamefont {Dresselhaus}\ \emph {et~al.}(2010)\citenamefont
  {Dresselhaus}, \citenamefont {Dresselhaus},\ and\ \citenamefont
  {Jorio}}]{Dresselhaus:2010aa}%
  \BibitemOpen
  \bibfield  {author} {\bibinfo {author} {\bibfnamefont {Mildred~S}\
  \bibnamefont {Dresselhaus}}, \bibinfo {author} {\bibfnamefont {Gene}\
  \bibnamefont {Dresselhaus}}, \ and\ \bibinfo {author} {\bibfnamefont {Ado}\
  \bibnamefont {Jorio}},\ }\href@noop {} {\emph {\bibinfo {title} {Group
  theory: application to the physics of condensed matter}}}\ (\bibinfo
  {publisher} {Springer},\ \bibinfo {year} {2010})\BibitemShut {NoStop}%
\bibitem [{\citenamefont {Koster}\ \emph {et~al.}(1963)\citenamefont {Koster},
  \citenamefont {Dimmock}, \citenamefont {Wheeler},\ and\ \citenamefont
  {Statz}}]{pointgroup}%
  \BibitemOpen
  \bibfield  {author} {\bibinfo {author} {\bibfnamefont {George~F}\
  \bibnamefont {Koster}}, \bibinfo {author} {\bibfnamefont {John}\ \bibnamefont
  {Dimmock}}, \bibinfo {author} {\bibfnamefont {Robert}\ \bibnamefont
  {Wheeler}}, \ and\ \bibinfo {author} {\bibfnamefont {Hermann}\ \bibnamefont
  {Statz}},\ }\href@noop {} {\emph {\bibinfo {title} {Properties of the
  thirty-two point groups}}}\ (\bibinfo  {publisher} {The MIT Press},\ \bibinfo
  {year} {1963})\BibitemShut {NoStop}%
\end{thebibliography}

%

\end{document}